\documentclass[twocolumn]{aa}
\usepackage{graphicx}
\usepackage{txfonts}
\usepackage{comment}
\usepackage{float}
\usepackage{subfig}
\usepackage[dvipsnames]{xcolor}
\usepackage[utf8]{inputenc}
\usepackage[T1]{fontenc}
\usepackage{amsfonts}
\usepackage[colorlinks=true,allcolors=blue]{hyperref}
\usepackage{mathrsfs}
\usepackage[version=4,arrows=pgf-filled,
textfontname=sffamily,
mathfontname=mathrm]{mhchem}

\begin{document} 
\title{One-dimensional and time-dependent modelling of complex organic molecules in protostars}
\author{Le Ngoc Tram\inst{1}\thanks{Corresponding author: Le Ngoc Tram \newline \email{nle@strw.leidenuniv.nl}},
Serena Viti\inst{1,2,3},
Katarzyna M. Dutkowska\inst{1}, 
Gijs Vermariën\inst{1,4},
Tobias Dijkhuis\inst{1,5},
Audrey Coutens\inst{6},
Timea Csengeri\inst{7},
Thiem Hoang\inst{8,9}
}

\institute{$^{1}$ Leiden Observatory, Leiden University, PO Box 9513, 2300 RA Leiden, The Netherlands\\
$^{2}$ Transdisciplinary Research Area (TRA) ‘Matter’/Argelander-Institut für Astronomie, University of Bonn, Bonn, Germany\\
$^{3}$ Department of Physics and Astronomy, University College London, Gower Street, London, UK \\
$^{4}$ SURF, Amsterdam, The Netherlands\\
$^{5}$ Leiden Institute of Chemistry, Leiden University, 2300 RA Leiden, The Netherlands\\
$^{6}$ Institut de Recherche en Astrophysique et Planétologie, Université de Toulouse, CNRS, CNES, 9 av. du Colonel Roche, 31028 Toulouse Cedex 4, France \\
$^{7}$ Laboratoire d’astrophysique de Bordeaux, Univ. Bordeaux, CNRS, B18N, allée Geoffroy Saint-Hilaire, 33615 Pessac, France \\
$^{8}$ Korea Astronomy and Space Science Institute, Daejeon 34055, Republic of Korea \\
$^{9}$ Korea University of Science and Technology, 217 Gajeong-ro, Yuseong-gu, Daejeon, 34113, Republic of Korea \\
} 
\date{Accepted December 22nd, 2025}
\titlerunning{Complex molecules with new chemical modelling UCLCHEM}
\authorrunning{Tram et al., 2025}
\abstract
{
Complex organic molecules (COMs), the building blocks of life, have been extensively detected under various physical conditions, from quiescent clouds to star-forming regions. They therefore serve as excellent tracers for the local physical and chemical properties of these environments. Proper models that are capable of grasping the formation and destruction of COMs are crucial to understanding observations. However, given that distinct COMs may be detected from different locations and at varying times, we improve UCLCHEM - a gas-grain chemical code - to a one-dimensional, time-dependent model, tailored to protostars. In this update, we examine two stages of a protostar: the prestellar and heating stages, incorporating a simple radiative mechanism for both the internal and external radiation fields of the cloud. This approach relies on the key assumption that the dust and gas temperatures are completely coupled. Ultimately, we implement an updated version of our model to interpret observations obtained through both single-dish and interferometry under varying conditions, including a SgrB2(N1) hot core, massive Galactic clumps and a hot core in Orion. We show that our model could reproduce these observations well, highlighting that some COMs are positioned at a higher temperature in the envelope, whereas others are from the lower temperature, potentially leading to misinterpretation when using a single-point model. In a particular case of SgrB2(N1), the best model indicates that the cosmic-ray ionisation rate significantly exceeds the value typically used for the standard interstellar medium. Our model shows as an efficient computational tool particularly useful for better insights into observations of COMs.
}

\maketitle
\section{Introduction}\label{sec:intro}

It is generally accepted that complex organic molecules (COMs, molecules with more than 6 atoms, cf. \citealt{2009ARA&A..47..427H}) are the building blocks of life. There are to date a plethora of observations of COMs towards hot cores associated with high-mass star-forming regions (see e.g. \citealt{1986ApJS...60..819C,2008A&A...482..179B,2008ApJ...672..352R,2019A&A...628A..10B}), and hot corinos associated with low-mass star-forming regions (see e.g. \citealt{2000A&A...357L...9C,2012ApJ...757L...4J,2016ApJ...830L...6J,2016A&A...590L...6C,2017MNRAS.469.2230M}). Initially, COMs are believed to form within the ice mantles on dust particles and later transferred to the gas phase when the dust grains are heated to 100-300$\,$K (see, e.g., \citealt{2008ApJ...682..283G}; \citealt{2016ApJ...830L...6J}). However, COMs have also been detected in dark clouds and prestellar cores, where the temperature is rather cold (see, e.g. \citealt{2014ApJ...795L...2V,2021A&A...649L...4A,2023ASPC..534..379C}), which challenges the previous scenario. Various non-thermal processes have been proposed to desorb COMs, including photodesorption, reactive desorption, cosmic ray-induced desorption, and sputtering (see \citealt{2014FaDi..168....9V,2023ASPC..534..379C} for comprehensive reviews). An alternative gas-phase formation route for some COMs has also been proposed (\citealt{2015MNRAS.453L..31B,2017MNRAS.468L...1S}).

Chemical modelling is often used to understand observations of COMs as well as to reveal the chemical properties of the observed regions. Among others, the astrochemical open-source code UCLCHEM\footnote{\href{https://uclchem.github.io}{https://uclchem.github.io}} has been routinely used to interpret molecular observations  (\citealt{2017AJ....154...38H}, \citealt{2025A&A...703A..46D}, Vermariën et al. in prep). This framework primarily uses a zero-dimensional and time-dependent model, incorporating the three-phase gas-grain chemical networks that include the gas-phase, surface, and bulk (everything below the surface). However, the existing UCLCHEM model restricts the capacity to identify the spatial distribution of the sources of COMs, which can introduce bias into the interpretation of the observations. In addition, interferometers provide remarkably high spatial resolution for observations of COMs, necessitating a model with higher spatial dimensionality. 

Therefore, in this work, we improve the one-dimensional treatment of radiation and visual extinction in the UCLCHEM framework, tailored to low- and high-mass protostars. Note that the UCLCHEM code could already simulate several gas cells (see, e.g. \citealt{1999MNRAS.305..755V,2004MNRAS.354.1141V,2010MNRAS.407.2511A}), while, however, maintaining the same values of the gas volume density and temperature for all cells, and disregarding the effect of the internal UV radiation field originated from the central source. Consequently, visual extinction and temperature are overestimated. Additionally, photon-related phenomena may remain uncorrected; this happens when only the external radiation field is taken into account; the UV photons may not reach depths from the surface towards the centre. However, incorporating an internal radiation field can facilitate the penetration of UV photons from the interior outward. These effects influence chemical processes and can introduce biases.
Our model physically follows the variations in the core's envelope concerning physical properties such as gas densities, visual extinction, and temperature, potentially reducing the bias in predicting the abundance of COMs.

The structure of this paper is as follows. In Section \ref{sec:modelling}, we describe the radial and time-dependent properties of the gas density, visual extinction, and dust temperature for our 1D model. The results in qualitative comparisons with observations in various sources, including the SgrB2 (N1) hot core, Galactic massive clumps, and the Orion hot core, are given in Section \ref{sec:applications}. The limitations of our model are discussed in Section \ref{sec:limitations}, and the conclusions of our work are given in Section \ref{sec:conclusions}.

\section{One-dimensional UCLCHEM modelling} \label{sec:modelling}
This section describes our improvement of the distance- and time-dependent profiles for gas density, visual extinction, and temperature in our model. 

\begin{figure}[!ht]
    \centering
    \includegraphics[width=0.8\linewidth]{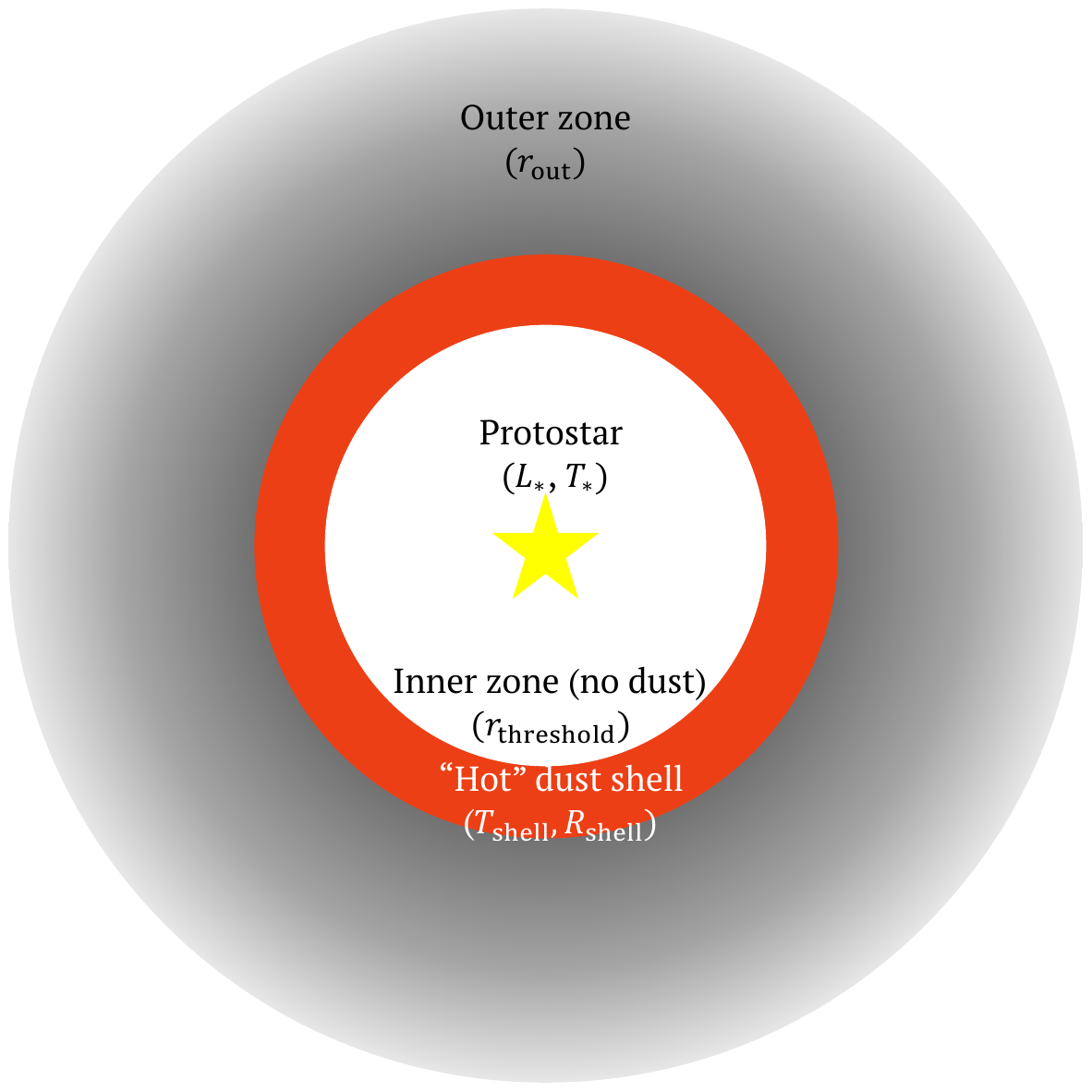}
    \caption{The sketch of the protostellar core adopted in this model, including a central source characterised by the bolometric luminosity ($L_{\ast}$) and the stellar temperature ($T_{\ast}$). The inner zone is characterised by the sublimation threshold of dust grains ($T_{\rm sub}$) and the threshold radius ($r_{\rm threshold}$). The thin shell of hot dust is determined by the shell temperature ($T_{\rm shell}$) and its radius ($R_{\rm shell}$). The outer region is given by the outer radius ($r_{\rm out}$).}
    \label{fig:sketch}
\end{figure}

We adhere to a conventional evolution of a protostar consisting of two distinct stages. The first is the collapsing (pre-stellar) phase, marked by material convergence due to gravitational instability, leading to a temporal rise in gas density and ultimately producing a density profile that follows a power-law decrease with radius. Consequently, various regions of the precursor cloud evolve at different rates; near the centre, because of the initially denser gas, this evolution proceeds more rapidly. The temperature is approximately constant through this stage because the radiation intensity from the surrounding medium is rapidly attenuated under such dense conditions. The next stage is the heating phase. During this stage, the physical phenomenon is the opposite. The formation of a protostar at the centre raises the temperature of the nearby gas and dust, and the gas density remains unvaried. Initially, the accretion process is the primary source of this heating, but as the protostar evolves, its luminosity takes precedence. Note that in proximity to the central luminous source, the dust particles are destroyed (no dust survives) if the temperature is higher than the sublimation temperature of the dust grain (e.g. $T_{\rm threshold} \sim 1500\,$K for silicate grains; \citealt{2015ApJ...806..255H}). When the temperature is below this threshold, there is a thin dust shell. This layer is irradiated by stellar radiation and plays an important role in heating the gas and dust in the outer zone. Figure \ref{fig:sketch} shows a sketch of the protostar scenario adopted in this work. Our assumption of the spherical protostellar core simplifies the source structure but we note that deviations from this model are supported by observations in the form of jets, outflows, and the disk. Such inclusions are beyond our scope of this work. 
\subsection{Radial physical profiles: Gas densities and visual extinctions}
We divide the radial profile into smaller $S$ segments with a constant radial spacing. Each gas parcel is determined by the distance $r$ to the centre. The gas volume density at a distance $r$ in the cores is adopted in the Bonnor-Ebert sphere (\citealt{1956MNRAS.116..351B}) as
\begin{equation}\label{eq:ngas}
    n_{\rm gas}(r)=\left\{
    \begin{array}{l l}
        n_{\rm 0} & \quad {\rm ~for~} r\le r_{\rm 0},\\
        n_{\rm 0}\left(\frac{r}{r_{\rm 0}}\right)^{-\alpha} & \quad {\rm ~for~} r>r_{\rm 0},
    \end{array}\right.
\end{equation}
where $n_{0}$ is the maximum density in the centre and is a parameter of our model, along with $\alpha$ and $r_{0}$. 

In a given gas cell, the gas column density from the centre to that cell ($N^{\rm centre2cell}_{\rm gas}$) and from the edge to that cell ($N^{\rm edge2cell}_{\rm gas}$) are
\begin{equation} \label{eq:Ngas_r}
    \begin{split}
     N^{\rm centre2cell}_{\rm gas}(r)=&\left\{
    \begin{array}{l l}
        n_{0}r & \quad {\rm ~for~} r\le r_{\rm 0},\\
         n_{0} r_{0} + \frac{n_{0}r_{0}}{\alpha-1}\left[1-\left(\frac{r}{r_{0}}\right)^{-\alpha+1}\right] & \quad {\rm ~for~} r>r_{\rm 0},
    \end{array}\right. \\
     N^{\rm edge2cell}_{\rm gas}(r)=&\left\{
    \begin{array}{l l}
         n_{0} r_{0} \left(\frac{\alpha}{\alpha-1}-\frac{r}{r_{0}}\right) & \quad {\rm ~for~} r<r_{\rm 0}, \\
        \frac{n_{0}r_{0}}{\alpha-1} \left(\frac{r}{r_{0}}\right)^{1-\alpha} & \quad {\rm ~for~} r> r_{\rm 0},\\
    \end{array}\right.
    \end{split}
\end{equation}

The corresponding visual extinctions are computed as
\begin{equation} \label{eq:Av_r}
    \begin{split}
        A^{\rm centre2cell}_{\rm V}(r) &=N^{\rm centre2cell}_{\rm gas} / 1.61\times 10^{21} ~~~ {\rm mag,} \\
        A^{\rm edge2cell}_{\rm V}(r) &=N^{\rm edge2cell}_{\rm gas} / 1.61\times 10^{21} ~~~ {\rm mag.}
    \end{split}
\end{equation}

\subsection{Radial physical profiles: radiation field and temperatures}
In order to consistently model the 1-dimensional effects of radiation, we first calculate the radiation from the protostars as well as the radiation from the thin dust shell and then derive the temperature.
\subsubsection{Stellar radiation}
The dimensionless energy density of the radiation at a distance $r$ from the centre is calculated as
\begin{equation} \label{eq:Ustar}
        U(T_{\ast}) = \frac{\int^{20\,\rm \mu m}_{0.091\,\rm \mu m} u_{\lambda}(T_{\ast})e^{-\tau_{\lambda}}d\lambda}{u_{\rm ISRF}}
\end{equation}
where $u_{\lambda}(T_{\ast})=L_{\lambda}(T_{\ast})/(4\pi r^{2}c)$ is the spectral energy density for no dust obscured, $u_{\rm ISRF}=8.64\times 10^{-13}\,{\rm erg\,cm^{-3}}$ is the energy density of the interstellar radiation field, and $\tau$ is the optical depth. We assume that the centre emits as a black body, which yields
\begin{equation}
    u_{\lambda}(T_{\ast})=\frac{4\pi R^{2}_{\ast} \pi B_{\lambda}(T_{\ast})}{4\pi r^{2}c}
\end{equation} 
while the optical depth is determined by the extinction curve as 
\begin{equation}
    \tau_{\lambda}=(A_{\lambda}/A_{\rm V})\times A^{\rm centre2cell}_{\rm V}/1.086
\end{equation}
with the extinction curve ($A_{\lambda}/A_{\rm V}$) being adopted from \cite{1989ApJ...345..245C}.
\begin{figure*}
    \sidecaption
        \includegraphics[width=12cm]{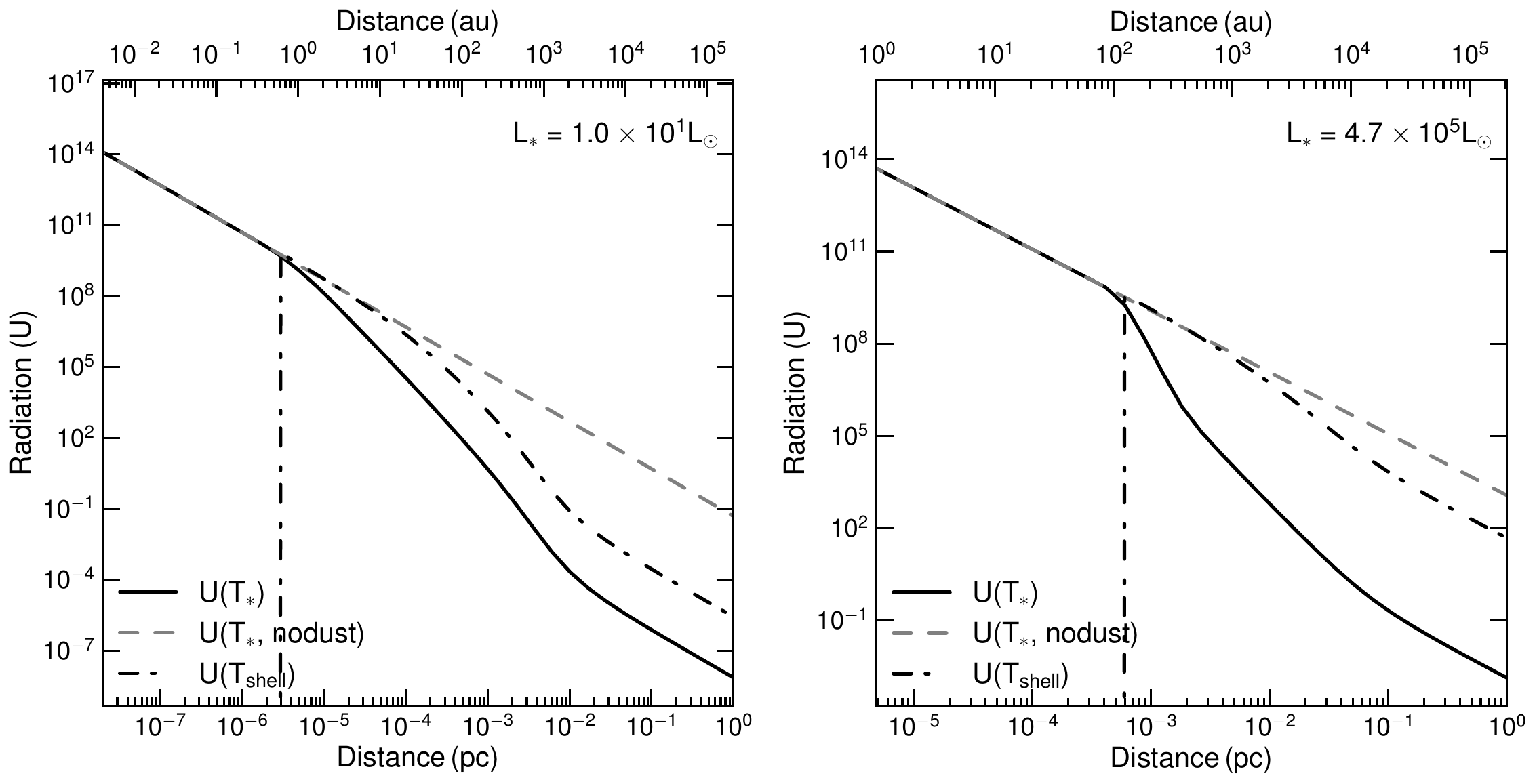}
        \caption{The variation of stellar and dust shell radiations with respect to radial distance for low-mass (left panel) and high-mass (right panel) protostellar cores. For $r<r_{\rm threshold}$ (marked by the vertical dashed-dotted line), the stellar radiation decreases as $U(T_{\ast},{\rm nodust}) \sim r^{-2}$.  For $r>r_{\rm threshold}$, this stellar radiation ($U(T_{\ast})$) is quickly reduced because of dust attenuation. The radiation in the outer region is mainly dominated by the emission from the dust shell ($U(T_{\rm shell})$).}
        \label{fig:radial_decompose}
\end{figure*}
\begin{figure*}
    \sidecaption
        \includegraphics[width=12cm]{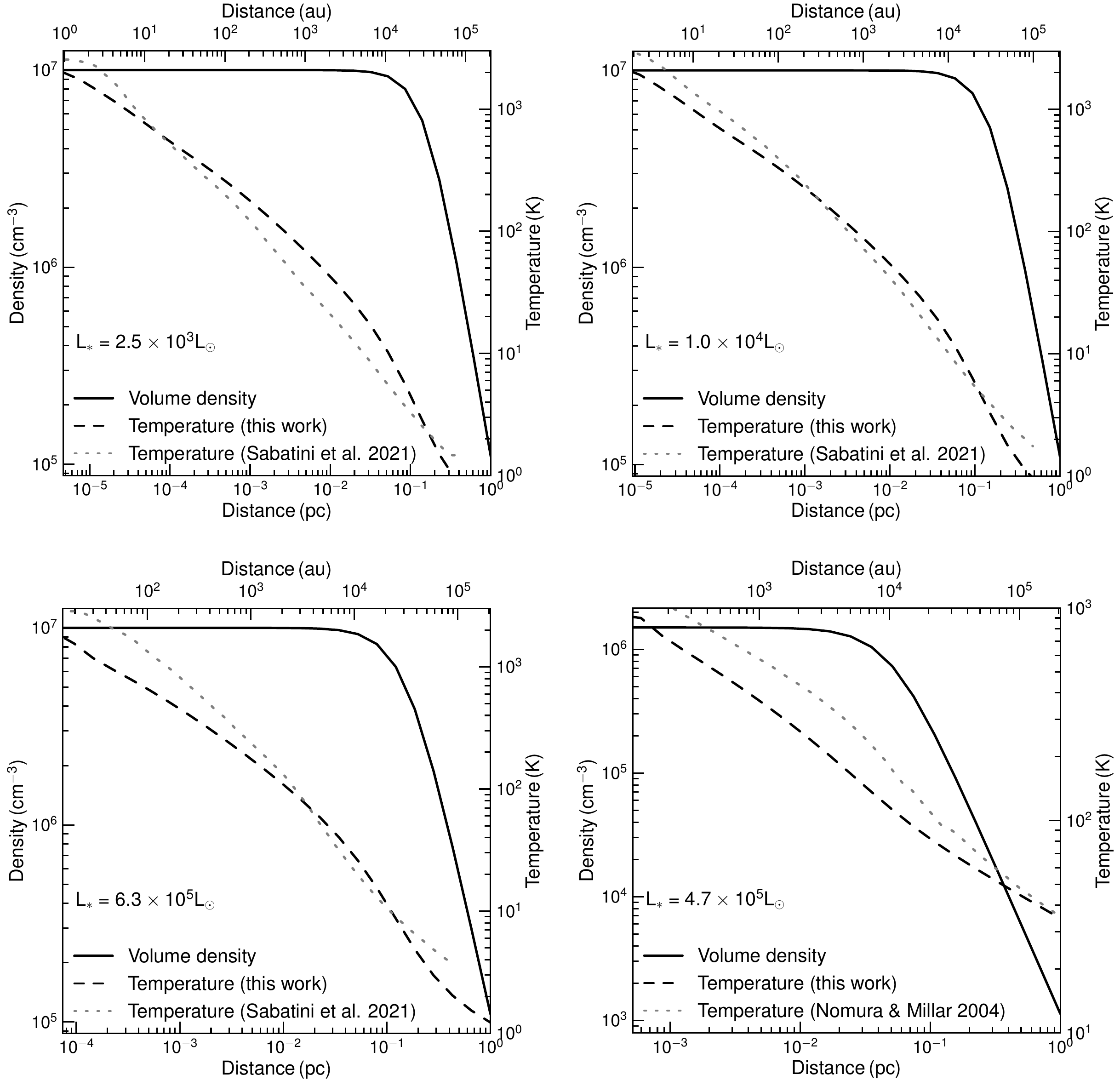}
        \caption{Examples of the radial profiles of the gas density (solid line, left y-axis) and dust temperature (dashed line, right y-axis) in protostars of different $L_{\ast}$ embedded in a molecular core for $r\geq r_{\rm threshold}$. In each panel, the our temperature profile is compared against the profiles found in literature. For the profiles in \cite{2021A&A...652A..71S}, we used $n_{0}=10^{7}\,\rm cm^{-3}$ and $r_{0}=0.15\,\rm pc$ as same as the authors. For the profile from \cite{2004A&A...414..409N}, we compare for the case of $n_{0}=1.5\times 10^{6}\,\rm cm^{-3}$, and $r_{0}=0.05\,\rm pc$.}
        \label{fig:phys_profiles}
\end{figure*}

The dust survival distance, $r_{\rm threshold}$, is intrinsically dependent on the threshold temperature ($T_{\rm threshold}$) of the grain material (above which the grains are destroyed by sublimation) and is approximated, as in \cite{2021ApJ...921...21H}, as\footnote{Our $T_{\rm threshold}$ is also known as $T_{\rm sub}$ in \cite{2021ApJ...921...21H} and the references therein.}
\begin{equation} \label{eq:rsub}
    r_{\rm threshold} \simeq 155.3\left(\frac{L_{\rm bol}}{10^{6}L_{\odot}}\right)^{0.5}\left(\frac{T_{\rm threshold}}{1500\,{\rm K}}\right)^{-5.6/2} ~~~{\rm au}
\end{equation}
with $T_{\rm threshold} = 1500\,$K as in \cite{2015ApJ...806..255H}. We observe that stellar radiation decreases as $r^{-2}$ for $r<r_{\rm threshold}$, but is more greatly attenuated due to the additional optical depth for $r\geq r_{\rm threshold}$. In this work, we focus solely on the positions where $r\geq r_{\rm threshold}$ (that is, dust can survive).

\subsubsection{Radiation from the dust shell}
The ``hot'' dust shell absorbs the stellar radiation from the source and re-emits as thermal emission farther out in the envelope. At a distance $r>r_{\rm threshold}$ in the outer region, the dimensionless radiation energy density attributed by this shell is
\begin{equation} \label{eq:Ushell}
    \begin{split}
    U(T_{\rm shell}) &= \frac{\int^{20\,\rm \mu m}_{0.091\,\rm \mu m} L_{\lambda}(T_{\rm shell})e^{-\tau_{\lambda}}d\lambda}{4\pi r^{2}c u_{\rm ISRF}} \\
    &= \frac{\int^{20\,\rm \mu m}_{0.091\,\rm \mu m} 4\pi R^{2}_{\rm shell} \pi B_{\lambda}(T_{\rm shell})e^{-\tau_{\lambda}}d\lambda}{4\pi r^{2}c u_{\rm ISRF}} 
    \end{split}
\end{equation}
where $T_{\rm shell}$ is calculated from $U(T_{\ast})(r=r_{\rm threshold})$ with an assumption that the hot shell is thin enough ($d_{\rm shell}\ll 1$) to prevent any radiative processes from occurring within it, resulting in the constant temperature of the shell ($T_{\rm shell}(R_{\rm shell}) = T_{\rm shell}(R_{\rm shell}+d_{\rm shell})$).

It is worth noting that the stellar radiation is responsible for the inner zone, while the dust shell at longer wavelengths is primarily responsible for heating in the outer regions (i.e. $U(T_{\ast})\ll U(T_{\rm shell})$ when $r\gg r_{\rm threshold}$). For example, a high-mass protostar with $T_{\ast}=4.5\times 10^{4}\,$K has a maximum emitted wavelength of approximately $65\,$nm, which is in the far ultraviolet range. The ultraviolet photons from the star are absorbed quickly after the dust shell. For lower-mass protostars, the peak of the radiation energy spectrum is shifted to a longer wavelength, and the photons can penetrate more deeply into the outer region, but the dust shell is still dominant for $r\gg 1$. These features are shown in Figure \ref{fig:radial_decompose} with the mean wavelengths for these two radiation fields decomposed in Figure \ref{fig:wave_decompose}.  

\subsubsection{Temperature profile}
If the dust is in equilibrium, which is a reasonable assumption in the case of large dust grains, the dust temperature is related to the radiation intensity as shown in \cite{2011piim.book.....D} as
\begin{equation} \label{eq:Tdust_r}
    \begin{split}
        T^{\rm sil}_{\rm d}(r) &= 16.4\,{\rm K}\times \left[U(T_{\ast}) + U(T_{\rm shell})\right]^{1/6} ~~~~~ {\rm silicate} \\
        T^{\rm car}_{\rm d}(r) &= 19.5\,{\rm K}\times \left[U(T_{\ast}) + U(T_{\rm shell})\right]^{1/5.6} ~~~ {\rm carbonaceous}
    \end{split}
\end{equation}
where $U(T_{\ast})$ and $U(T_{\rm shell})$ are the functions of distance and calculated from Eqs. \ref{eq:Ustar} and \ref{eq:Ushell}. The dust temperature is taken as the average as $T_{\rm d}(r) = \left[0.625\times \left(T^{\rm sil}_{\rm d}\right)^{4} + 0.375\times \left(T^{\rm car}_{\rm d}\right)^{4}\right]^{1/4}$ with silicate and carbonaceous referring to the dust composition. Figure \ref{fig:phys_profiles} shows the radial profiles for both the gas density (solid line on the left y-axis) and the dust temperature (dashed line on the right y-axis) corresponding to four different protostars with a luminosity of $L_{\ast} \sim 10^{3} - 10^{6}L_{\odot}$. The gas density remains uniform up to $r_{0}$ and decreases beyond this point. The dust temperature sharply reduces towards the surface as a result of the attenuation of the radiation field. These values of gas density and dust temperature at each radius are specified for the final values (when $t\rightarrow \infty$) in the time-dependent evolution at this position. As a benchmark, we compared the dust temperature profiles in our model with those in \cite{2021A&A...652A..71S} and \cite{2004A&A...414..409N} for the same central luminosity sources and found consistent temperature profiles. The temperature values at the very far edges of the envelope are for comparison with the literature. In practice, we stop at $T\sim 10-20\,$K as shown in Table \ref{tab:input_params}.

\subsection{Time-dependent physical profiles: pre-stellar stage}
For a given gas parcel with boundary location $r_{i}$, we model the change in physical properties over time as follows. In the pre-stellar collapse phase, the gas volume density increases from an initial $n_{0}(t=0,r)$ to $n_{\rm gas}(t\rightarrow \infty, r)$ described in Eq. \ref{eq:ngas} as
\begin{equation} \label{eq:ngas_t}
    \begin{split}
        \frac{dn(t,r_{i})}{dt} = &b_{c}\left(\frac{n(t,r_{i})^{4}}{n(t=0,r_{i})}\right)^{1/3} \\ 
        &\times \left[24\pi G m_{\rm H} n(t=0,r_{i})\left(\left(\frac{n(t,r_{i})}{n(t=0,r_{i})}\right)^{1/2}-1\right)\right]^{1/2}
    \end{split}
\end{equation}
with $b_{c}=1$ the freefall factor, $G$ the gravitational constant, and $m_{\rm H}$ representing the mass of hydrogen nuclei. The collapse continues until the gas reaches the final density $n_{\rm gas}$ aimed at a specific gas parcel. To mimic the fact that the initial gas volume density is denser toward the centre, we calculate the term $n(t=0,r)$ from Eq. \ref{eq:ngas} with $n_{0}=2.18\times 10^{4}\,\rm cm^{-3}$ and $r_{0}=1.4\times10^{4}\,\rm au$ (see Eqs. 7 and 8 in \citealt{2018AJ....156...51P}).    

For each gas parcel, the column density is evaluated at every time step using the expression
\begin{equation}
    \Delta N_{\rm gas}(r_{i},t) = \frac{r_{\rm out}}{\rm S}n_{\rm gas}(r_{i},t)
\end{equation}
The cumulative column density extending from the cloud's edge to that parcel is then given by
\begin{equation}
    N^{\rm edge2cell}_{\rm gas}(r,t) = \sum^{r}_{r_{i}=r_{\rm out}}\Delta N_{\rm gas}(r_{i},t)
\end{equation}
which allows for calculating the visual extinction ($A^{\rm edge2cell}_{\rm V}(r,t)$) at each time step as well, using Eq. \ref{eq:Av_r}.

It is important to highlight that within our approximation, when \(t\) approaches infinity, we have \(\lim _{t\rightarrow \infty}n_{\rm gas}(r,t) \rightarrow n_{\rm gas}(r)\), as defined in Eq. \ref{eq:ngas}. Similarly, \(\lim_{t\rightarrow \infty} N_{\rm gas}(r,t) \rightarrow N^{\rm edge2cell}_{\rm gas}(r)\), as defined in Eq. \ref{eq:Ngas_r}, and \(\lim_{t\rightarrow \infty} A_{\rm V}(r,t) \rightarrow A_{\rm V}(r)\), as defined in Eq. \ref{eq:Av_r}.

\subsection{Time-dependent physical profiles: heating stage}
We turn on the central luminosity source (the protostar) after the pre-stellar stage. The gas and dust are heated by different mechanisms at different times, such as accretion, photospheric luminosity from gravitational contraction, and deuterium burning in the central star. As we do not model the evolution of the central luminosity, we assume that at a point $r$, the evolution of gas/dust temperature from $T_{\rm d}(t=0,r)$ to $T_{\rm d}(t\rightarrow \infty,r)$ (described in Eq. \ref{eq:Tdust_r}) is characterised by an empirical relation as in \cite{2010MNRAS.407.2511A} as 
\begin{equation} \label{eq:Tdust_t}
    T_{\rm d}(t,r) = T_{\rm d,0} + At^{B}\left(\frac{r}{r_{\rm out}}\right)^{-\beta} ~~~{\rm K}
\end{equation}
with the constants $A$ and $B$, defining the curvature of $T_{\rm d}$ vs. time, $T_{\rm d,0}=10\,$K the temperature of the ambient gas. Figure \ref{fig:phys_profiles} shows that $T_{\rm d} \sim r^{-0.5}$, and we then adopt $\beta=0.5$. The values of $A$ and $B$ are derived from combinations of
\begin{itemize}
    \item $T_{\rm d}(t=0,r)=10\,$K
\end{itemize}
and $T_{\rm d}(r)$ in Eq. \ref{eq:Tdust_r} with
\begin{itemize}
    \item $T_{\rm d}(r)=100\,$K after $10^{5}\,$yr for $1\times M_{\odot}$ star (\citealt{2010MNRAS.407.2511A}) or
    \item $T_{\rm d}(r)=100\,$K after $10^{5}\,$yr for $5\times M_{\odot}$ star or
    \item $T_{\rm d}(r)=300\,$K after $2\times 10^{5}\,$yr for $10\times M_{\odot}$ star or
    \item $T_{\rm d}(r)=200\,$K after $1.1\times 10^{5}\,$yr for $15\times M_{\odot}$ star (\citealt{1999MNRAS.305..755V, 2004MNRAS.354.1141V}) or
    \item $T_{\rm d}(r)=200\,$K after $7\times 10^{4}\,$yr for $25\times M_{\odot}$ star (\citealt{1999MNRAS.305..755V}) or
    \item $T_{\rm d}(r)=200\,$K after $2.8\times 10^{4}\,$yr for $60\times M_{\odot}$ star (\citealt{1999MNRAS.305..755V})
\end{itemize}

During this stage, the additional attenuation from the central source is characterised by the visual extinction from the centre to the parcel as $A^{\rm centre2cell}_{\rm V}(r,t) \equiv A^{\rm centre2cell}_{\rm V}(r)$ as computed in Eq. \ref{eq:Av_r}, due to the constant gas volume density.

\subsection{UV radiation fields}
In the heating stage, we add the UV radiation field from the central source (internal UV field), in addition to the external $G^{\rm ext}_{0}$ that is a parameter in our model. This internal field solely depends on the luminosity in the UV bands ($L_{\rm UV}$) of the source and is defined at the dust shell as
\begin{equation}
    G^{\rm int}_{0} = \frac{L_{\rm UV,\ast}}{4\pi c r^{2}_{\rm threshold} 5.29\times 10^{-14}\,{\rm erg\,cm^{-3}}}
\end{equation}
where $L_{\rm UV,\ast}$ is the integration of the luminosity of the central source, approximately a black body from 0.6 to 13.6 eV, which yields $L_{\rm UV,\ast} = 0.4-0.5\times L_{\ast}$. For a given position at a given time, this internal UV field is attenuated as $G^{\rm int}_{0}e^{-A^{\rm centre2cell}_{\rm V}}$.

\subsection{Chemical network}
 
In the version described in this study, the reaction network comprises 356 species and 8766 gas-phase reactions sourced from the latest UMIST22 database (\citealt{2024A&A...682A.109M}). The grain reactions are taken from \cite{2017AJ....154...38H} and \cite{2018MNRAS.474.2796Q}. The species in the chemical network are denoted as, for example, $\ce{H2O}$ (gas phase water), $\#\ce{H2O}$ (surface water), $@\ce{H2O}$ (bulk water), and $\$\ce{H2O}$ (ice water), with ice being the sum of surface and bulk.

Our model contains both thermal and nonthermal mechanisms to desorb the species from the surface into the gas phase. The latter processes include ultraviolet (UV) radiation, molecular hydrogen formation, direct cosmic rays, cosmic-ray-induced UV, and chemical desorption. It also includes the competition of Langmuir-Hinshelwood reactions by diffusion and desorption as described by \citet{Chang2007}. We refer the reader to \cite{2018MNRAS.474.2796Q} and \cite{2025A&A...703A..46D} for a complete description of the chemical network used.

UCLCHEM starts with gas species in atomic or ionic forms (see Table \ref{tab:initial_abund} for the total abundances related to hydrogen of elementary species) without molecules. For hydrogen, half is atomic and the other half is molecular. During the collapse phase, chemical reactions occur, leading to the formation of molecules within the gas phase and ice mantles on dust particles. The end abundances of this stage serve as the initial conditions for the subsequent heating stage.

\begin{table}[!ht]
    \centering
    \caption{Total initial elemental abundances with respect to the total number of Hydrogen nuclei (H), which are either atomic or ionic forms, except for H, where half is atomic and the other half is molecular.}
    \begin{tabular}{llllll}
        \hline
         Species & & & & & Abundances  \\
         \hline
         Helium & & & & & 0.1\\
         Carbon & & & & & 1.77$\times 10^{-4}$\\
         Oxygen & & & & & 2.34$\times 10^{-4}$\\
         Nitrogen & & & & & 6.18$\times 10^{-5}$\\
         Sulphur & & & & & 3.51$\times 10^{-6}$\\
         Magnesium & & & & & 2.26$\times 10^{-6}$\\
         Silicate & & & & & 1.78$\times 10^{-6}$ \\
         Iron & & & & & 2.01$\times 10^{-7}$ \\
         \hline
    \end{tabular}
    \label{tab:initial_abund}
\end{table}

\section{Results and applications} \label{sec:applications}
\subsection{Distance- and time-dependent physical properties of a molecular core}
\begin{table}[]
    \centering
    \caption{The range of radial distance from the centre for low-mass and high-mass protostars.}
    \begin{tabular}{c c c c c}
        \hline 
        \hline
        Central Luminosity & $r_{\rm min}$ & $r_{\rm max}$ & $T(r_{\rm min})$ & $T(r_{\rm max})$   \\
        (input) & (input) & (input) & (output) & (output) \\
        ($L_{\odot}$) & (pc) & (pc) & (K) & (K) \\
        \hline
        $10^{6}$ & 0.005 & 0.5 & 18 & 336 \\
        $10^{5}$ & 0.002 & 0.2 & 18 & 366 \\
        $10^{4}$ & 0.001 & 0.1 & 18 & 355 \\
        $10^{3}$ & $6\times 10^{-4}$ & 0.06 & 20 & 300 \\
        $10^{2}$ & $4\times 10^{-4}$ & 0.03 & 10 & 203 \\
        10       & $1.5\times 10^{-4}$ & 0.015 & 10   & 200    \\
        1        & $10^{-4}$ & 0.01 & 8   & 160     \\
        \hline 
        \hline
    \end{tabular}
    \begin{flushleft}
    \tablefoot{The central mass is based on the Hertzsprung-Russel diagram. All grids are computed with 100 gas cells. For $L_{\ast}>10^{2}L_{\odot
    }$, $r_{0}=0.05\,$pc and $n_{0}=10^{7}\,\rm cm^{-3}$ are adopted. For $L_{\ast}\leq 10^{2}L_{\odot}$, $r_{0}=0.005\,pc$ and $n_{0}=10^{8}\,\rm cm^{-3}$ are used. Grain size $a=0.5\,\mu$m is adopted to account for the grain growth effect.}
    \end{flushleft}
    \label{tab:input_params}
    \caption{Key physical parameters for the modelling along the west- and south-directions of the SgrB2(N1) hot core.}
    \begin{tabular}{c|c|c}
        \hline
        \hline
        parameters & SrgB2(N1)-West & SrgB2(N1)-South \\
        \hline
         $L_{\ast}$ ($L_{\odot}$) & $6\times 10^{6}$& $6\times 10^{6}$ \\
         $n_{0}$ ($\rm cm^{-3}$) & $10^{7}$ & $10^{7}$ \\
         $r_{0}$ (pc) & 0.02 & 0.012 \\
         $r_{\rm out}$ (pc) & 0.3 & 0.3 \\
         $\zeta_{\rm scale}$ & 1-100 & 1-100 \\
         $G_{0}$ & $10^{3}$ & $10^{3}$ \\
         \hline
         \hline
    \end{tabular}
    \begin{flushleft}
    \tablefoot{The luminosity of the SgrB2(N1) is adopted from \cite{2019A&A...628A...6S}, whereas the gas volume density is taken from \cite{2017A&A...604A..60B}. $r_{0}$ is an adjustable parameter; these two values are based on the radial profile of $X(\ce{CH3OH})$. Based on the Hertzsprung–Russell diagram, the mass of the central protostellar core is adopted as 60$\,M_{\odot}$.
   }
    \end{flushleft}
    \label{tab:sgrb2n1}
\end{table}
\begin{figure}[!ht]
    \centering
    \includegraphics[width=0.9\linewidth]{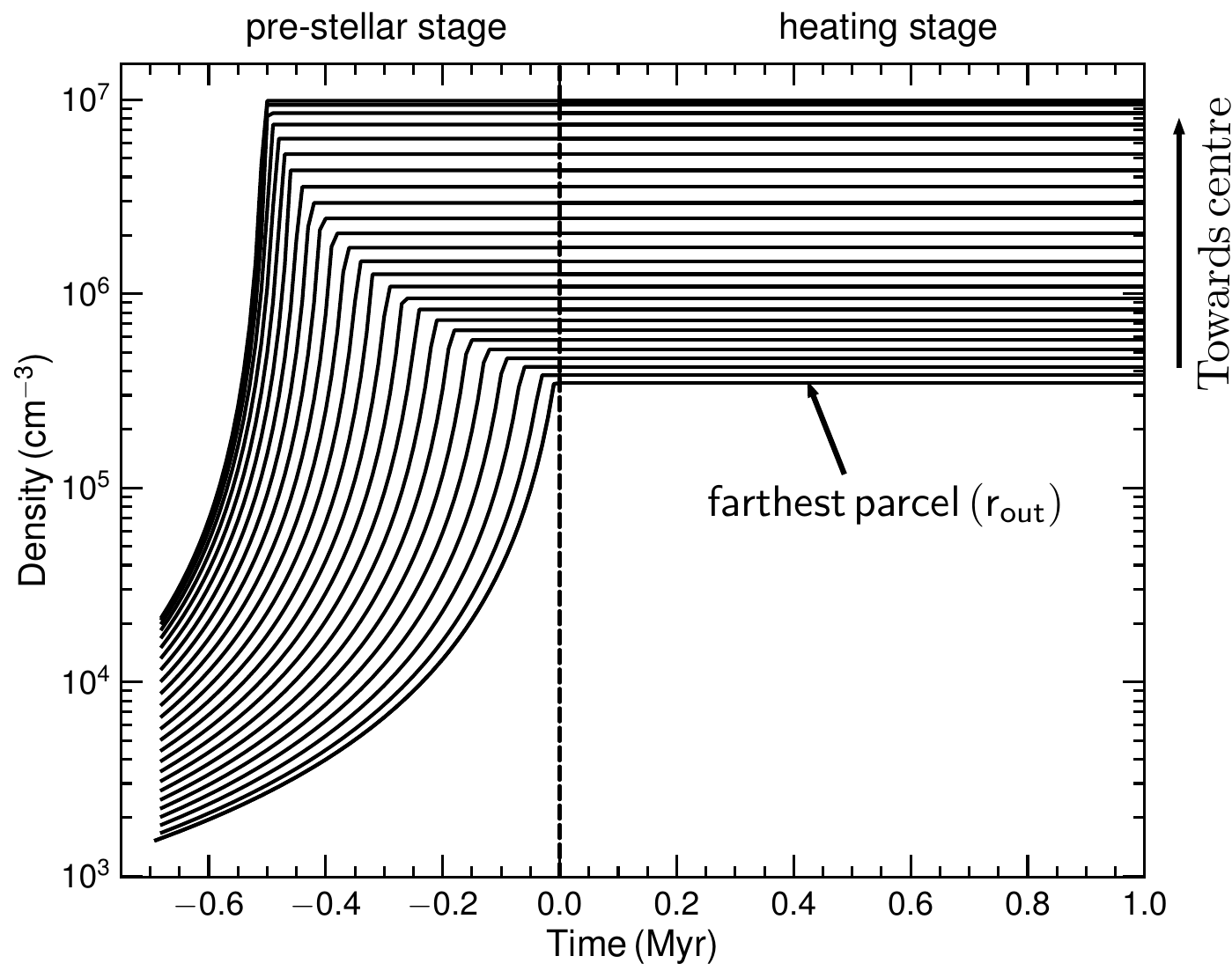} \\
    \includegraphics[width=0.9\linewidth]{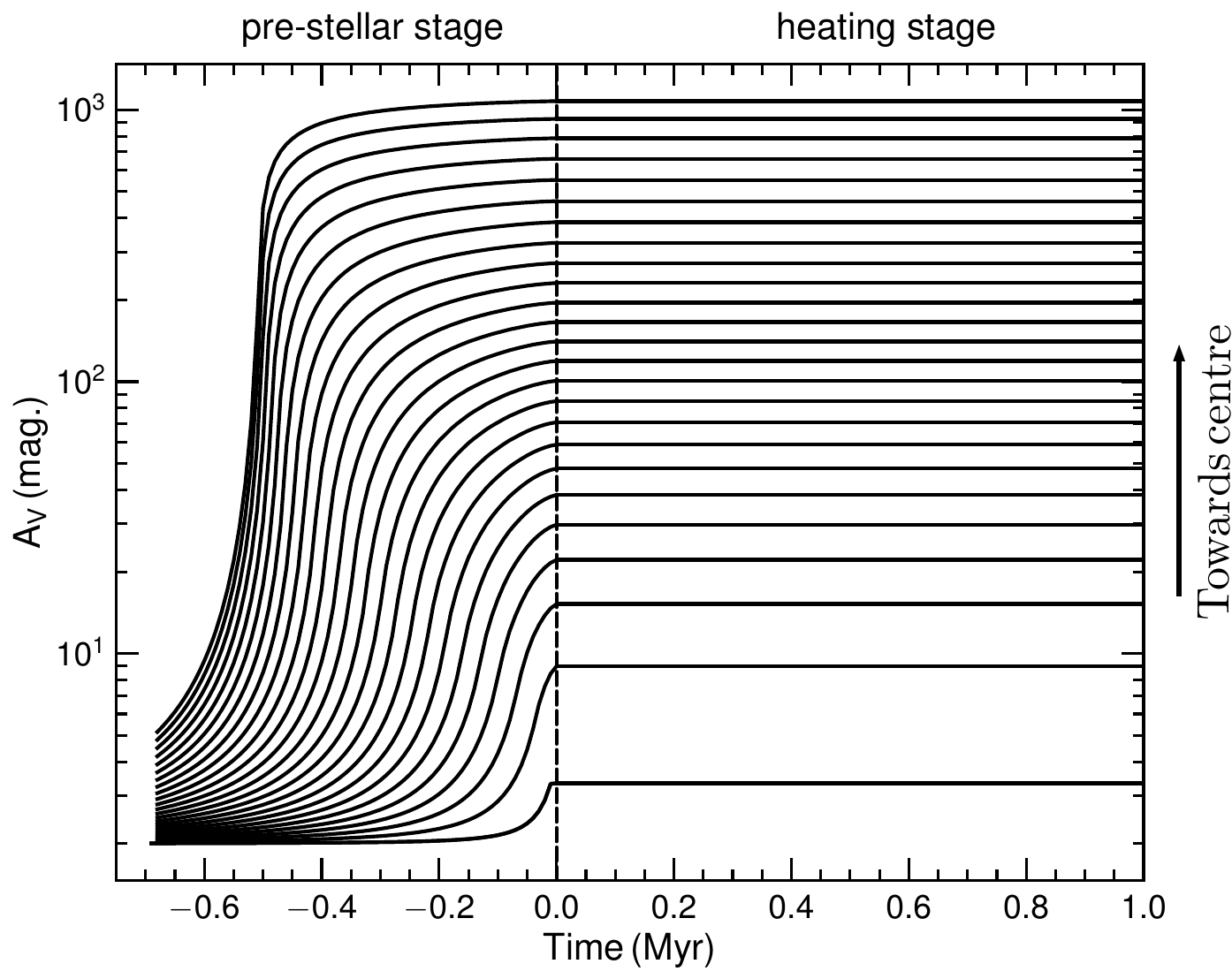} \\
    \includegraphics[width=0.9\linewidth]{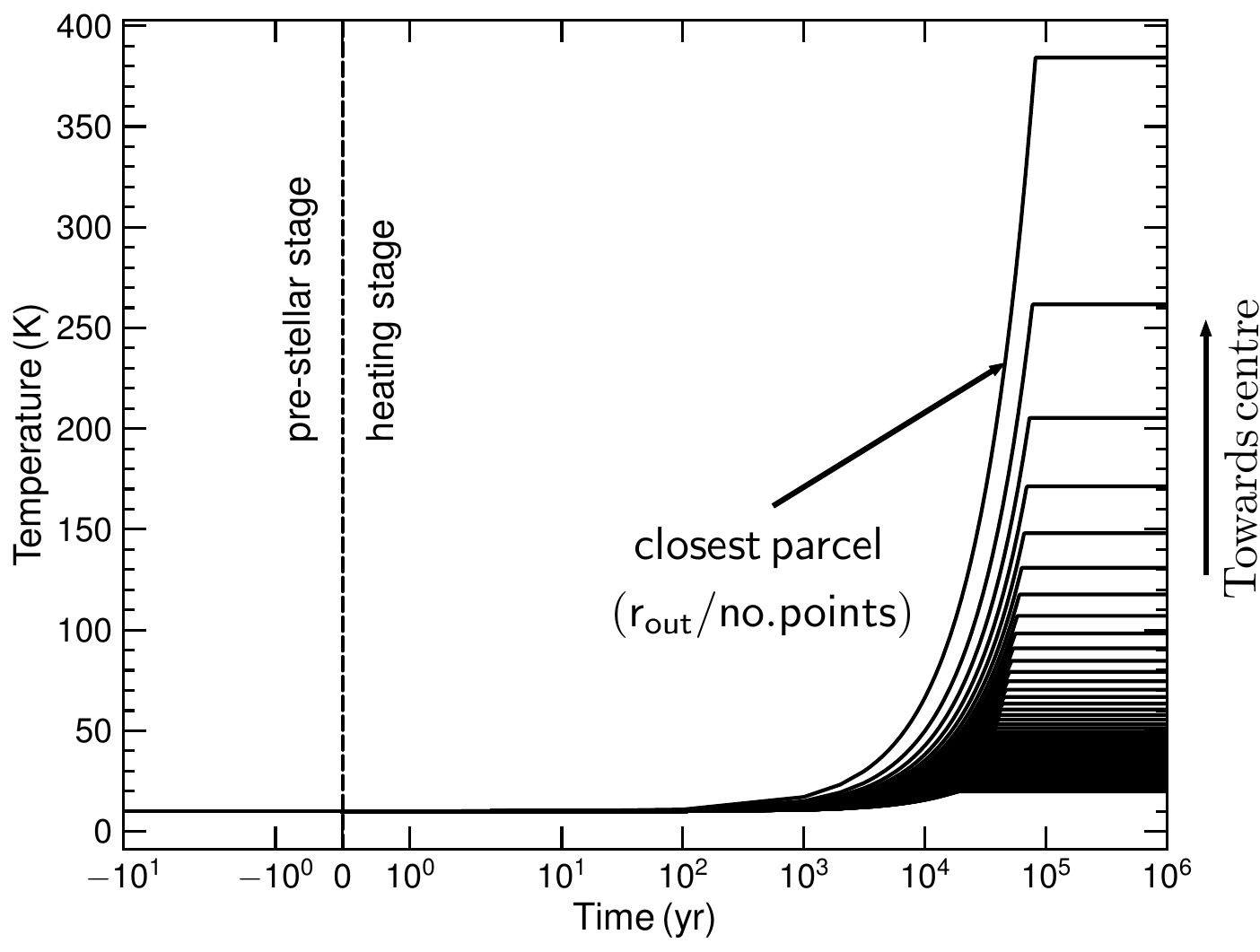}
    \caption{Examples of the profiles of the gas density (top), visual extinction from the surface (middle) and temperature (bottom). The negative time is for the pre-stellar phase, time of zero is where the protostar is formed. For the illustration, 50 over 100 gas parcels in the sampling are shown. The values of $n_{\rm gas}(r)$ estimated by Eq. \ref{eq:ngas} and $T_{\rm d}(r)$ estimated by Eq. \ref{eq:Tdust_r} is the final value of $n_{\rm gas}(t)$ in Eq. \ref{eq:ngas_t} and $T_{\rm d}(t)$ in Eq. \ref{eq:Tdust_t} at a given location $r$. These profiles are with $L_{\ast}=10^{5}\,L_{\odot}$, $n_{\rm in}=10^{7}\,\rm cm^{-3}$, $r_{\rm flat}=0.05\,\rm pc$ and $a=0.5\,\mu$m.}
    \label{fig:gasdense_gastemp}
\end{figure}

\begin{figure*}
    \centering
    \includegraphics[width=0.4\linewidth]{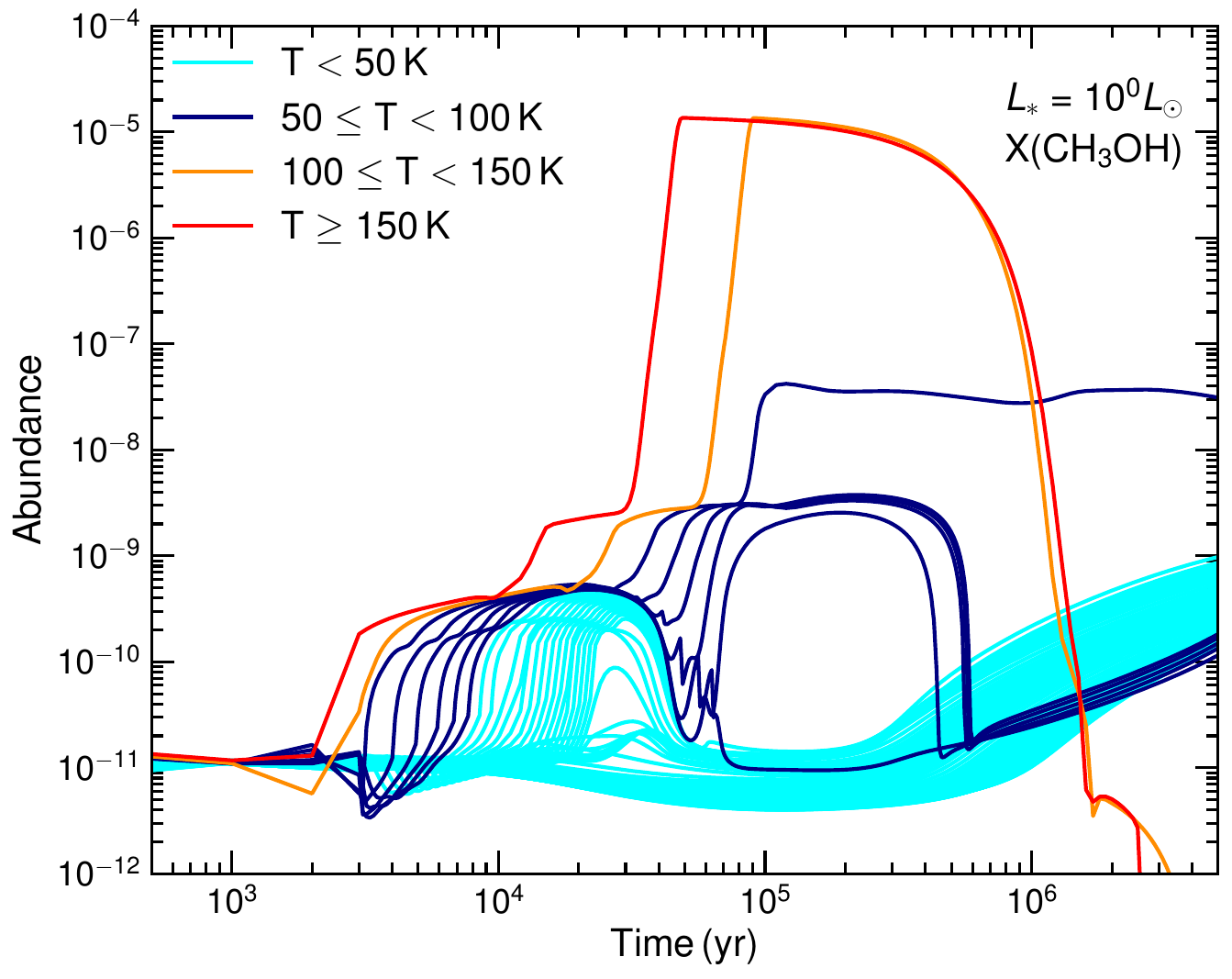}
    \includegraphics[width=0.4\linewidth]{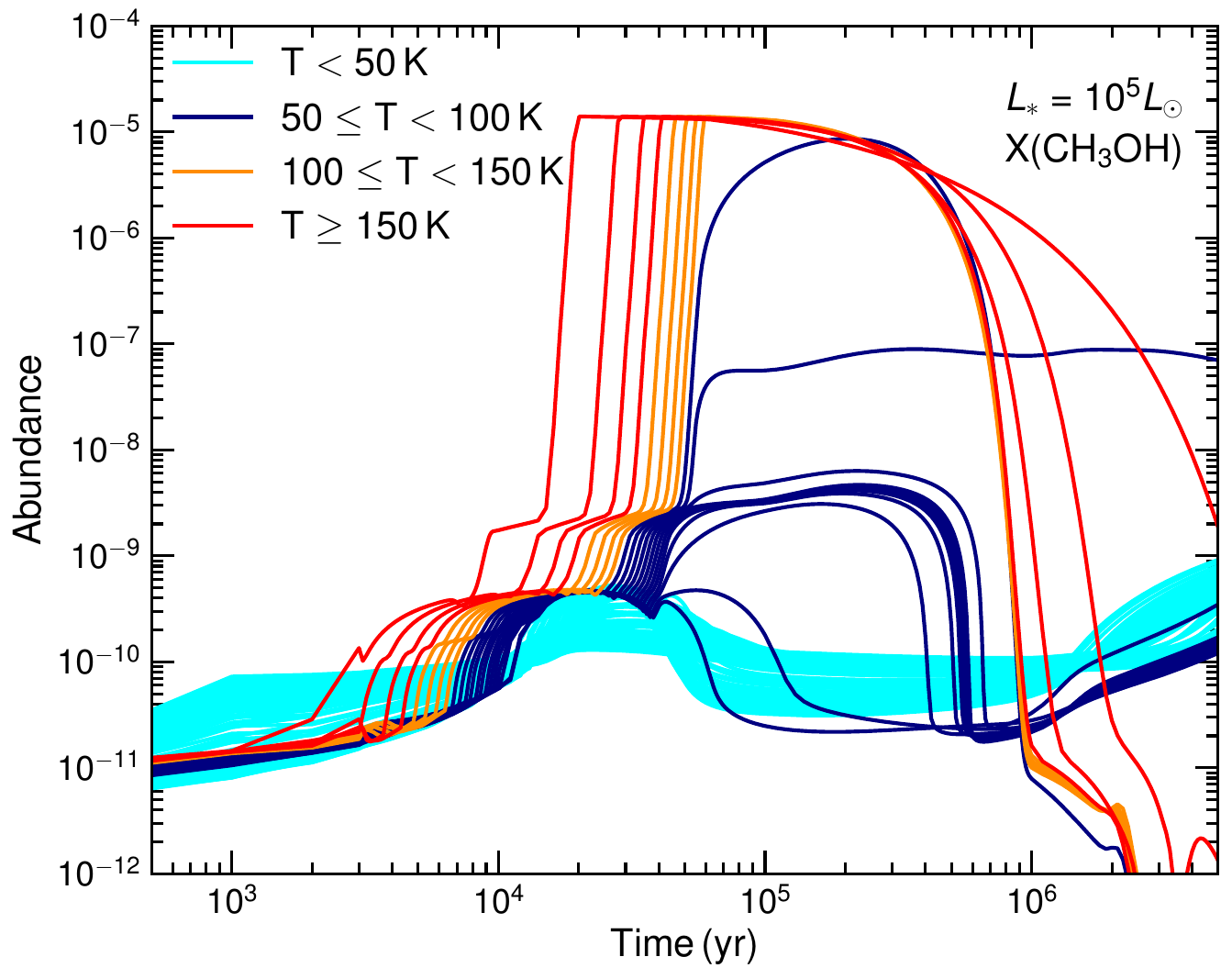}
    \caption{Example of the variation of $\ce{CH3OH}$ with time for $L_{\ast}=10^{0}\,L_{\odot}$ (left) and $L_{\ast}=10^{5}\,L_{\odot}$ (right). Different lines are for 100 parcels (full grid) from the center. The lines are colored with the corresponding values of temperature at $t=10^{6}\,$yr. The abundance of $\ce{CH3OH}$ decreases sharply at late time because of the reaction with neutral and ion species. For instance, at $t=5\times 10^{5}\,\rm yr$, the most important destruction reactions are $\ce{H3O+} + \ce{CH3OH} \rightarrow \ce{CH3OH2+} + \ce{H2O}$ and $\ce{H3+} + \ce{CH3OH} \rightarrow \ce{CH3+} + \ce{H2O} + \ce{H2}$.}
    \label{fig:Xch3oh}
    \includegraphics[width=0.8\linewidth]{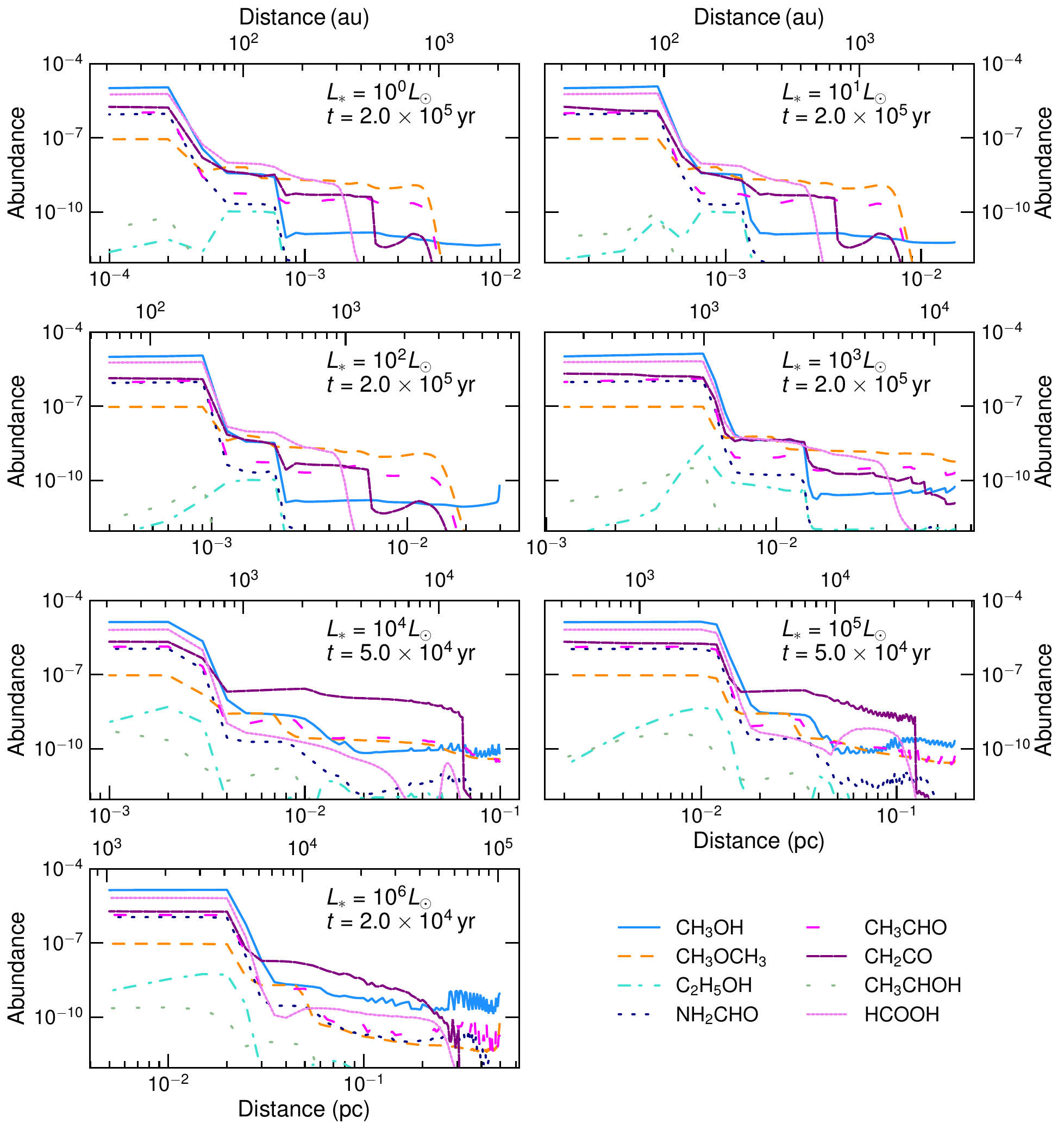}
    \caption{The abundance of various COMs as a function of distance based on different source luminosities at different ages. While the chosen age is arbitrary, it is consistent with the fact that a more massive protostar evolves faster than the less massive one. Typically, the abundance remains unchanged near the centre and decreases farther out. In the case of a more massive protostellar core, the abundance extends over a larger volume.}
    \label{fig:X_distance}
\end{figure*}
\begin{figure*}
    \centering
    \includegraphics[width=0.46\linewidth]{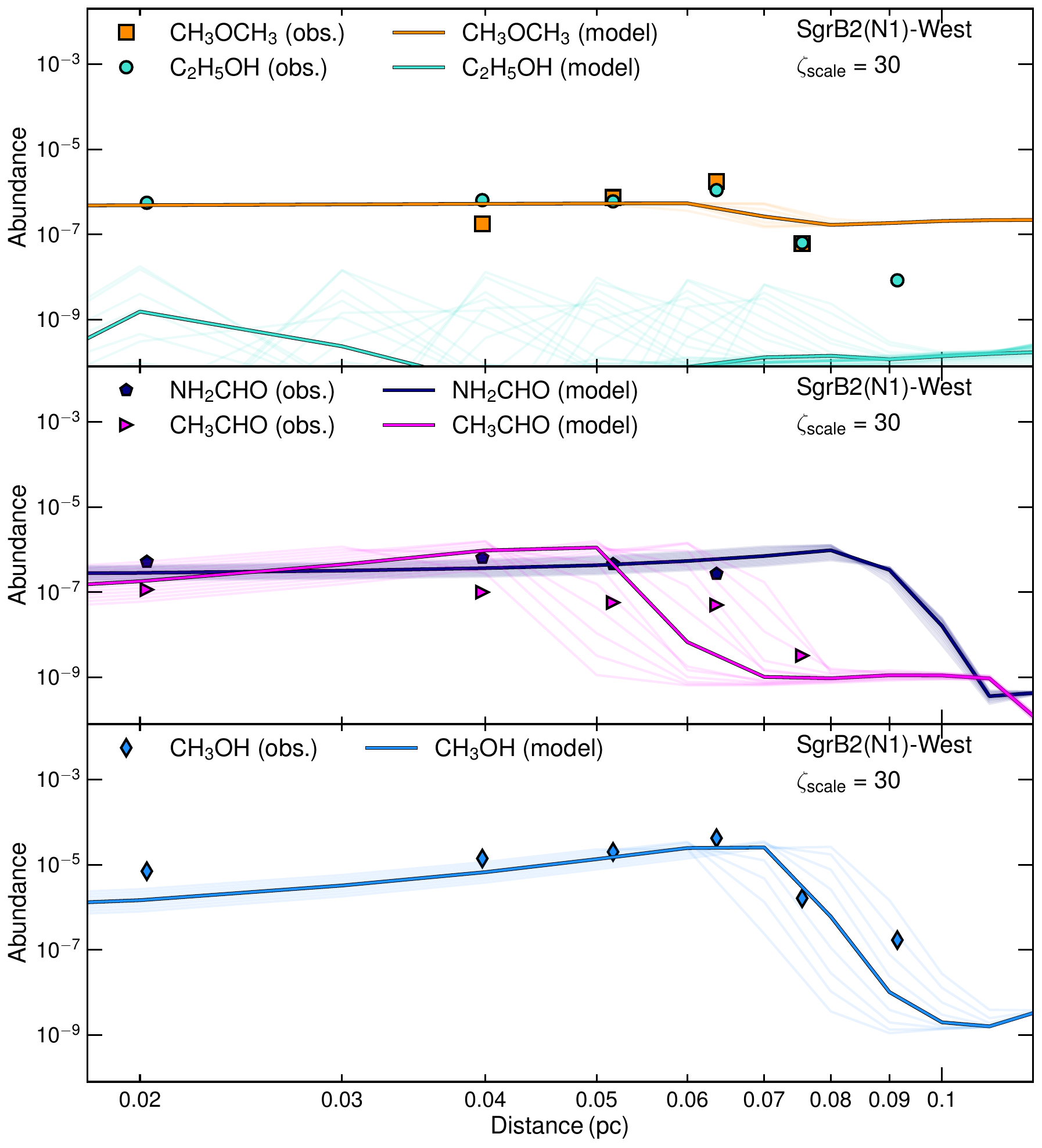}
    \includegraphics[width=0.46\linewidth]{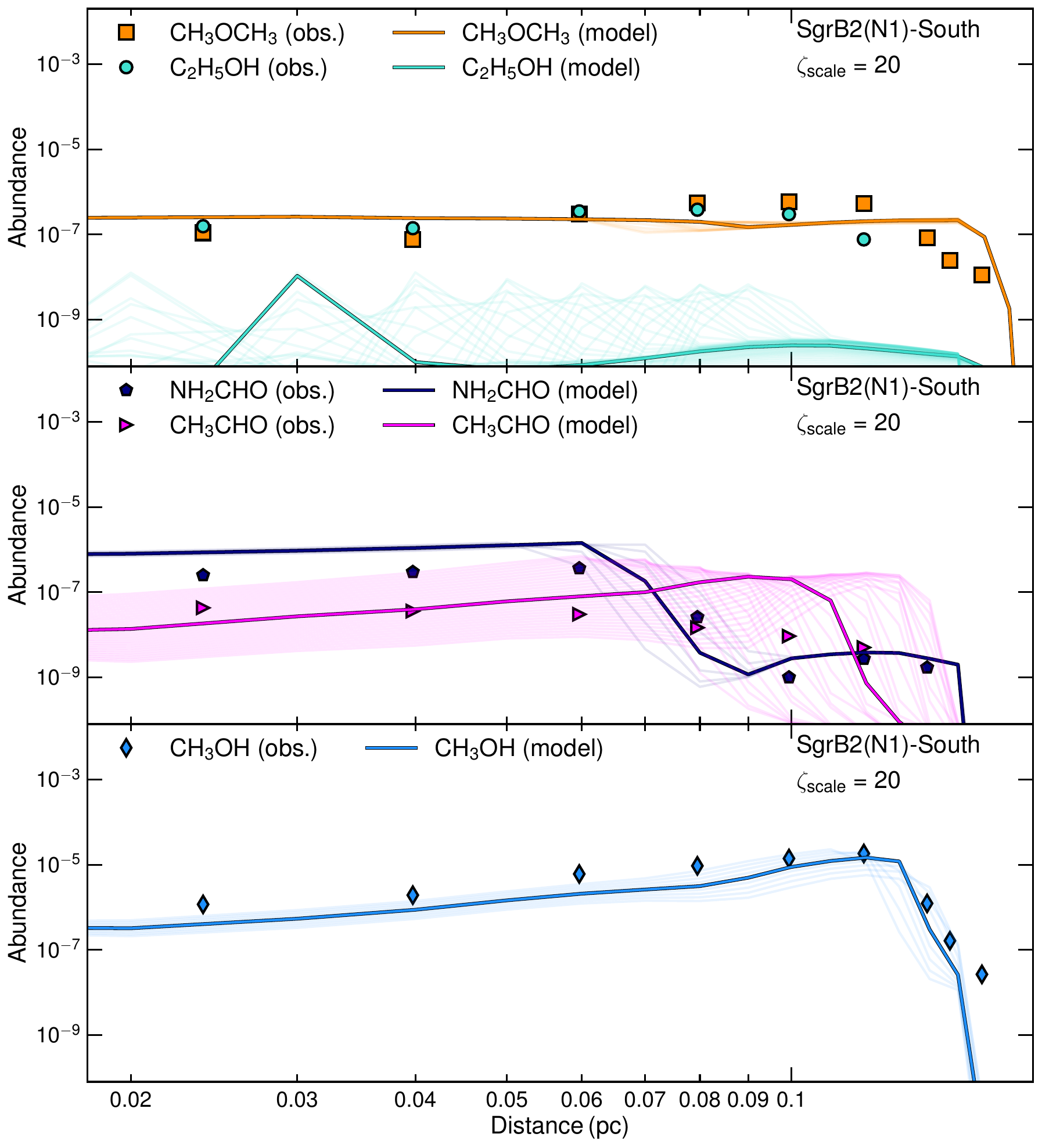}
    \caption{The radial profiles of some COMs as a function of radial distance moving westward (left panel) and southward (right panel) from the SgrB2(N1) hot core with $G_{0}=10^{3}$. The symbols represent observational data, while the prominent solid lines illustrate the best models (the lighter lines denote models corresponding to the ages within $\pm 1\sigma$ around the age determined by the best $\chi^{2}$). Our model can nicely reproduce the radial profiles of $\ce{CH3OH}$, $\ce{CH3OCH3}$, $\ce{NH2CHO}$ and $\ce{CH3CHO}$, but fails to explain the $\ce{C2H5OH}$ profile. To better match observations for all COMs, the cosmic ray ionisation rate is required to be higher than the typical value of ISM. For the westward, $\zeta_{\rm scale}=30$ corresponds to $\zeta (0.2\,\rm pc)\simeq 60\times \zeta_{\rm ISRF}$, while $\zeta_{\rm scale}=20$ corresponds to $\zeta (0.2\,\rm pc)\simeq 50\times \zeta_{\rm ISRF}$ for the southward. The best fit period correspond to $(3-7)\times 10^{4}\,$yr.}
    \label{fig:sgrb2n1}
\end{figure*}
Figure \ref{fig:gasdense_gastemp} shows an example of the time-dependent physical profiles for different gas parcels estimated from our model. The top panel shows that parcels closer to the centre, in the prestellar phase, experience a faster evolution of their gas volume density because their initial $n(t=0)$ is higher. When they reach the final value assigned to their positions (estimated by Eq. \ref{eq:ngas}), they remain constant.

The middle panel shows the corresponding profiles of the visual extinction from the edge inwards. Similarly to the gas density profile, the gas parcel deeper inside the core quickly becomes obscured as the visual extinction increases to at least two orders of magnitude higher than the edge.

The bottom panel illustrates the time-dependent profile of the temperature. During the pre-stellar stage, this profile remains constant at 10$\,$K. During the heating stage, as the central source heats up, the gas temperature rises to its final value at the specified location depicted in Figure \ref{fig:phys_profiles}. Due to attenuation, the gas parcels closer to the centre experience a more rapid temperature increase compared to those situated farther out in the cloud.

\subsection{Distance- and time-dependent chemical properties of a molecular core}
Figure \ref{fig:Xch3oh} shows an example of the variation of the gas phase $\ce{CH3OH}$ over time for 100 gas parcels from the centre for two particular cases of $L_{\odot}$ (left panel) and $10^{5}\,L_{\odot}$ (right panel). Two distinct features are seen: the abrupt increase in the abundance in proximity to the central source is due to the thermal sublimation. In the envelope, where the temperature is below 100$\,$K, the abundance of $\ce{CH3OH}$ is rather low due to the majority of molecules being frozen on the grains. Non-thermal desorption processes desorb a limited amount of $\ce{CH3OH}$ molecule. For gas parcels with the highest temperatures, the abundances of $\ce{CH3OH}$ and other complex molecules are seen to rapidly decrease at late times; our chemical analysis shows that the reactions of neutral ions drive this destruction. For example, the most destructive reaction to destroy $\ce{CH3OH}$ at $5\times 10^{5}\,$yr is $\ce{H3O+} + \ce{CH3OH} \rightarrow \ce{CH3OH2+} + \ce{H2O}$ with a rate of $-4.17\times 10^{-19}$ $\rm cm^{-3}\,s^{-1}$.

Figure \ref{fig:X_distance} illustrates how the gas phase abundance of selected COMs changes with the radial distance for low-mass and high-mass protostars. For methanol, Figure \ref{fig:Xch3oh} shows that the abundance reaches its peak and remains uniform near the core, subsequently decreasing as it extends towards the envelope. A more luminous central source results in a larger region where the abundance is constant. The input parameters are listed in Table \ref{tab:input_params}.

\begin{figure*}[!ht]
    \centering
    \includegraphics[width=0.9\textwidth]{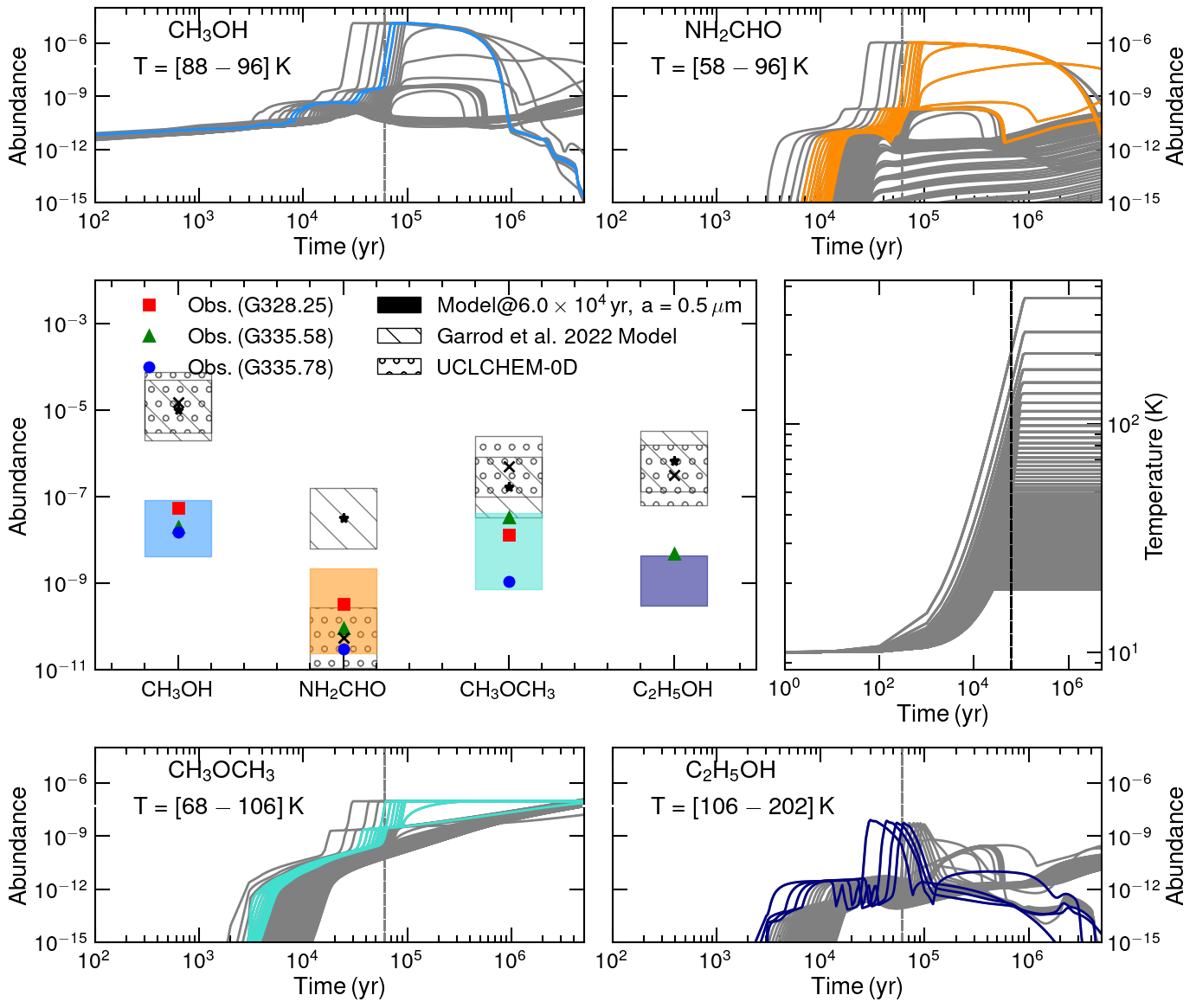}
    \caption{Comparison between model and observations of G328.25, G335.58 and G335.78. The middle left panel shows the model's lower and upper bounds that best cover the observations at a given time of $6\times 10^{4}\,$yr with a grain size of 0.5$\,\mu$m. These areas of best coverage relate to various locations in the envelope, marked by colored lines on the top and bottom panels. In these sub-panels, different lines are for different locations within the cloud. The different colour lines correspond to the different values for the temperatures. The middle right panel shows the temperature profile for all locations in our model. Comparisons with the 0D model (UCLCHEM-0D with $G_{0}$=1 and \citealt{2022ApJS..259....1G}) are also provided (mean prediction values are shown by the 'x'/asterisk, while the hatched areas are for a factor of 5 scattered around these mean values).}
    \label{fig:Galactic_massive_clumps_shaded}
\end{figure*}
\begin{figure*}
    \centering
    \includegraphics[width=0.9\linewidth]{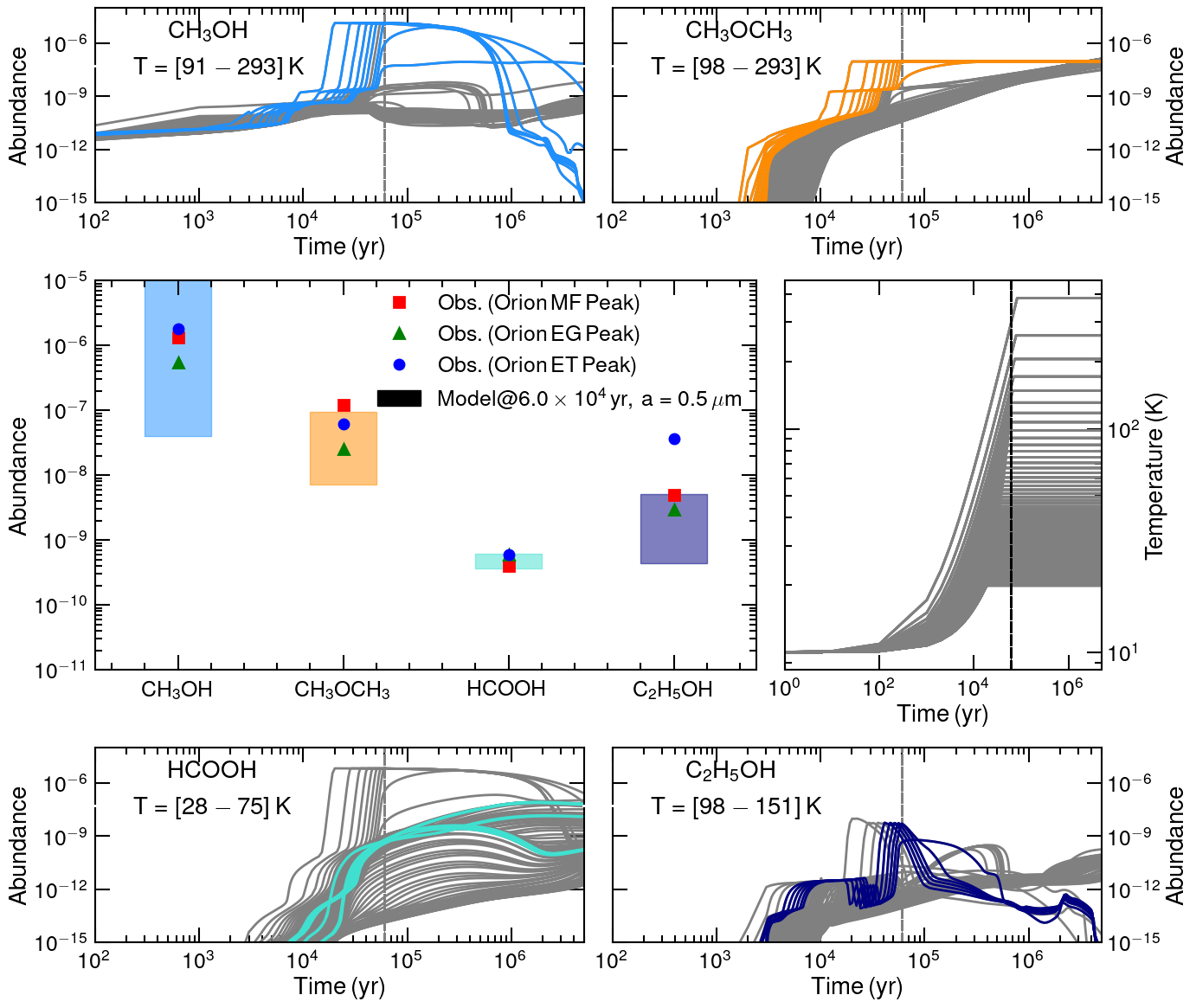}
    \caption{Similar to Figure \ref{fig:Galactic_massive_clumps_shaded} but with observations of Orion hot core for $\ce{CH3OH}$, $\ce{CH3OCH3}$, $\ce{HCOOH}$ and $\ce{C2H5OH}$ towards MF, EG and ET peaks at a specific cloud age of $6\times 10^{4}\,$yr. Our model underestimates $X(\ce{C2H5OH})$ observed in the ET peak.}
    \label{fig:Orion}
\end{figure*}
\subsection{Application: COMs abundances in the SgrB2 (N1) hot core} \label{sec:SgrB2}
In this section, we compare the predictions of our model with the spatial distribution of abundances of some COMs reported by \cite{2022A&A...665...A96} within the ReMoCA (Re-exploring Molecular Complexity with ALMA) survey (\citealt{2019A&A...628A..10B}). This survey used ALMA to observe towards the hot core N1 in the giant molecular cloud Sagittarius B2-North (hereafter SgrB2(N1)). The observations are made with different setups, covering an angular resolution from $0.75''$ in setup 1 to $0.3''$ in setup 5 (we refer to Table 2 in \citealt{2019A&A...628A..10B} for more details). The abundances are taken in two directions: west and south to the central hot core.

We modelled the radial abundances of some oxygen-bearing COMs, including $\ce{CH3OH}$, $\ce{CH3OCH3}$, $\ce{C2H5OH}$, $\ce{NH2CHO}$ and $\ce{CH3CHO}$ with the main input parameters described in Table \ref{tab:sgrb2n1}. The luminosity and the mass of the central source are fixed, while we adjust the parameter $r_{0}$ because it regulates the gas volume density profile and thus regulates the radiative process. The best values of $r_{0}$ are based on a comparison of the radial profile of $X(\ce{CH3OH})$ from the model and observations. Finally, we vary the cosmic rate ionisation rate from the standard value in the ISM ($\zeta_{\rm ISM}=1.3\times 10^{-17}\,\rm s^{-1}$) with a scale factor ($\zeta_{\rm scale}=[1, 100]$); then the cosmic-ray ionisation rate at each location and time step in our model is $\zeta/\zeta_{\rm ISM}\times \zeta_{\rm scale}$ where the cosmic-ray attenuation towards the centre is considered following \cite{2022ApJ...934...63O} as $\log_{10}\zeta = \sum_{k=0}^{9}c_{k}\log^{k}_{10}(N^{\rm edge2cell}_{\rm gas})$ with the coefficient $c_{k}$ given in their Table A1 (model $\mathscr{L}$). The external UV radiation field is adopted as $10^{3}$ times higher than the typical interstellar radiation field with $G_{0,\rm ISM}=1$ (see, e.g., \citealt{2004ApJ...600..214G}). 

For each COM, we compare the predicted profiles with observations over time to determine the best fits of the model and time, using the minimum of $\chi^{2}$ as
\begin{equation}
    \chi^{2} = \sum_{i=1}^{N} \left[\frac{\ln(X_{\rm model}) - \ln(X_{\rm obs,i})}{\ln(X_{\rm obs,i})}\right]^{2}
\end{equation}
The summation is taken for all data points $N$, and the predicted $X_{\rm model}$ is interpolated at the corresponding distance as in observations. The left panel of Figure \ref{fig:sgrb2n1} shows the comparison of the radial profile of the COMs predicted from our model with observations in the west direction toward the central hot core SgrB2(N1). We note that our model can reproduce these profiles reasonably well for a higher cosmic-ray ionisation rate with $\zeta_{\rm scale} \simeq 30$, corresponding to the value of $\zeta(0.2\,\rm pc) \simeq 60\times \zeta_{\rm ISM}$ (see Figure \ref{fig:zeta_sgrb2n1} for the radial profile of $\zeta$). Nonetheless, our model cannot explain the observed profile of $\ce{C2H5OH}$ ($X_{\rm obs}(\ce{C2H5OH})\simeq 10^{-7}$). The explicit explanation is beyond the scope of this work and requires more extensive works. However, we discuss this issue in Section \ref{sec:GMP}.

Similarly, the right panel of Figure \ref{fig:sgrb2n1} shows the comparison for the south direction. Along this direction, $\ce{CH3OH}$ begins to decrease further from the centre ($>0.1\,$pc) compared to its decline along the west direction. As in the previous case, models with higher cosmic-ray ionisation rate are required ($\zeta(0.2\,\rm pc) \simeq 50\times \zeta_{\rm ISM}$ with the best $\zeta_{\rm scale}=20$, see Figure \ref{fig:zeta_sgrb2n1}) to reproduce all detections, but the prediction for the abundance of \ce{C2H5OH} is again too low compared to observations. The cosmic ray ionisation rates derived from our models for the hot core SgrB2(N1) are consistent with those of \cite{2019A&A...628...A27}, in which the authors show that $\zeta \geq 50\times \zeta_{\rm ISM}$ using the observations of COMs in the nearby hot cores SgrB2(N2-N5). 

It is important to note that the abundance decreases more rapidly to the west from the centre compared to the south. Furthermore, the value of $r_{\rm flat}$ is higher in the southern direction, suggesting a higher gas density in the south than in the west, consistent with Figure 8 in \cite{2022A&A...665...A96} and with the lower value of $\zeta$ predicted by our model in the south due to attenuation.

\subsection{Application: COMs abundances in massive protostars} \label{sec:GMP}
In this section, we assess our model's predictions against APEX observations of three massive protostars in our Galaxy: G328.25, G335.58, and G335.78 reported in \cite{2024A&A...686A.252B}.  The observations were made using different receivers of the APEX, whose beam size ranged from $18''$ to $36''$, and the abundances relative to the gas column density were scaled to the size of the observed emission area (cf. Section 5.6 in \citealt{2024A&A...686A.252B}). The luminosities of these sources are approximately the same as $10^{4}L_{\odot}$ as noted in Table 1 in \cite{2024A&A...686A.252B}, the differences in the observed COMs are not influenced by the radiative characteristics of these sources, which presents a perfect scenario for exploring other effects, such as the physical conditions of the sources. In this work, we particularly examine some specific oxygen-bearing COMs, including $\ce{CH3OH}$, $\ce{NH2CHO}$, $\ce{CH3OCH3}$, $\ce{C2H5OH}$, for which \cite{2024A&A...686A.252B} demonstrated that the single-point model in \cite{2022ApJS..259....1G} overestimates the abundances of these COMs (see their Figure 13). The authors suggested that the overestimation may arise due to the fact that the sources are in a stage where nonthermal sublimation predominantly determines abundances, whereas the thermal sublimation area is compact.

Figure \ref{fig:Galactic_massive_clumps_shaded} shows the lower and upper limits of our models (the shaded area in the largest panel) that cover the observed abundance of each species at a given time (the limits, both lower and upper, are determined by the distances $r_{1}$ and $r_{2}$ of the model where the abundances $X(r_{1})<X({\rm obs})<X(r_{2})$). This plot is for particular time values of $7\times 10^{4}\,\rm yr$ and a grain size of 0.5$\,\mu$m (the effect of the grain size is shown in Figure \ref{fig:Galactic_massive_clumps}). The height of these areas is due to the model's grid at that given time. From these best models, we determine the gas parcels from which the COMs are detected, and the temperature ranges highlighted by the colour lines in the corresponding upper and lower panels. Our temperature range is found to be roughly consistent with the excitation temperatures for these COMs in \cite{2024A&A...686A.252B} (see their Figure 11). 

We compare our best models with the single-point (0D) model of the original UCLCHEM with a standard UV field ($G_{0}=1$) and the one presented in \cite{2022ApJS..259....1G}\footnote{The predicted abundances are adopted from Table 17 in \citealt{2022ApJS..259....1G} (indicated by hatched areas)}. The 0D models tend to overestimate the abundance of the considered COMs, except for $X(\ce{NH2CHO})$, whereas our 1D model matches more closely with the observational data. This discrepancy arises because these COMs originate from different regions, necessitating the physical structure to be considered. For example, for temperatures exceeding 100\,K (thermal desorption dominates), our models predict abundance ($X(\ce{CH3OH})$, $X(\ce{CH3OCH3})$ and $X(\ce{C2H5OH})$) similar to those of the \citealt{2022ApJS..259....1G} model. It is important to mention that, unlike \cite{2022ApJS..259....1G}, our model does not incorporate non-diffusive chemistry. 

Interestingly, the 0D model can predict a higher $X(\ce{C2H5OH})$, which can better explain the observations in the SgrB2(N1) hot core mentioned in Section \ref{sec:SgrB2}. Therefore, we performed a series of tests with the 0D model to understand why our 1D model could not predict the observed  high abundance. We found that COMs become significantly less abundant when using higher values of $G_{0}$ (see Figure \ref{fig:Galactic_massive_clumps_shaded_G0=10}). In our 1D model, the high abundance of $\ce{C2H5OH}$ near the centre is affected by the internal UV field generated by the central source rather than by the external UV field. Therefore, the low abundance of $\ce{C2H5OH}$ may be an indication that in our 1D model either  the UV field profile is not correct or that we lack key chemical reactions for this species, such as non-diffusive reactions. If it is due to the former one, it remains unclear why only $\ce{C2H5OH}$ is too sensitive.

\subsection{Application: COMs abundances in the Orion hot core}
In this section, we compare our model with ALMA observations toward the methyl formate (MF) peak, the ethylene glycol (EG) peak, and the ethanol (ET) peak in the nearby hot core in the Orion BN/KL region as reported in \cite{2018A&A...620L...6T} at a resolution of $\sim 1.5''$. In this case, the central source luminosity is about $10^{5}L_{\odot}$. Figure \ref{fig:Orion} presents a comparison of the specific abundances of $\ce{CH3OH}$, $\ce{CH3OCH3}$, $\ce{HCOOH}$, and $\ce{C2H5OH}$ as predicted by our model (represented by shaded areas in the largest subfigure) at a given specific cloud age and those observed (indicated by coloured symbols; taken from Table 2 in \citealt{2018A&A...620L...6T}) in Orion BN/KL. In general, our model is effective in explaining the observed abundances and shows that the emitting regions of these COMs change with the cloud's age (colour lines in the figures on the top and bottom). The best coverage shows that $\ce{CH3OH}$, $\ce{CH3OCH3}$, and $\ce{C2H5OH}$ originate at higher temperatures compared to $\ce{HCOOH}$, and our model cannot reproduce $X(\ce{C2H5OH} \sim 3.5\times 10^{-8})$ at the Orion ET peak. The potential explanation might be an excessively strong UV field or the absence of chemical reactions involving $\ce{C2H5OH}$ as discussed in Section \ref{sec:GMP}.

\section{Limitations} \label{sec:limitations}
There are three main sources of uncertainty in our model, which include the decoupling in the temperature of the dust and the gas, the beam-average and the structure of the envelope, and we discuss each of these caveats in the following.

First, our model does not incorporate the heating and cooling processes, which might affect the temperature distributions of gas and dust and subsequently result in different COM abundances as discussed in this study. The gas-dust coupling, which is expected to hold during the heating stage, may in fact not hold at lower densities (e.g. \citealt{2019ApJ...884..176I} shows that $T_{\rm dust}\simeq T_{\rm gas}$ for $n_{\rm H}\geq 10^{6}\,\rm cm^{-3}$); therefore, the impact of the heating and cooling processes will primarily be significant in the collapse phase. To explicitly understand how these effects could impact predicted abundances of COMs, our upcoming work will incorporate the decoupling of the temperatures of the dust and gas. 

Second, the predicted quantities of COMs from our 1D model do not incorporate the beam average, which introduces bias when compared to observations, particularly those from single-dish observations. This factor necessitates employing a model of a higher dimension than our current 1D approach, since it is essential to accurately compute the COMs density across all sight lines within a spatial region before convolving into the intended beam size. Note that along each line of sight, the physical profiles of density and temperature vary differently from the radial profile adopted from the 1D model. 

Third, the physical characteristics of the protostellar cores do not account for the presence of clumpy structures or outflows or variations in luminosity. We expect the temperature profile to be more extended in the clumpy envelope, since the radiative transfer processes differently from the smoothed continuous envelope in this work. We notice that our model can be further improved by adapting the time-dependent luminosity of the protostellar cores (e.g., see Figure 3a in \citealt{2015A&A...575A..68C}).

\section{Conclusions} \label{sec:conclusions} 
In this study, we update the 0D-UCLCHEM chemical modelling of protostars to a simple and computationally efficient one-dimensional, time-dependent model. We then compare the predictions of our novel model with existing observations from a range of astronomical targets. Specifically, we conducted these comparisons for the SgrB2(N1) hot core (Figure \ref{fig:sgrb2n1}), infrared quiet galactic massive clumps (G328.25, G335.58, G335.78) (Figure \ref{fig:Galactic_massive_clumps_shaded}), and the hot core in Orion BN/KL (Figure \ref{fig:Orion}). We summarise our findings below: 
\begin{itemize}
    \item[-] We considered the formation of a protostar as happening in two phases: collapse and heating phases. Initially, during the collapse stage, the gas volume density increases over time as the temperature remains fixed. In the subsequent heating phase, a luminous source at the core is activated (controlled by the bolometric luminosity), leading to a temperature increase without altering the density. Initially, the gas is denser near the core, while the dust temperature is estimated using a basic radiative transfer setup, assuming an equilibrium between the radiation field and temperature.
    
    \item[-] Comparison with the SgrB2(N1) hot core: we used the ALMA observations within the ReMOCA survey, reported in \cite{2022A&A...665...A96} for $\ce{CH3OH}$, $\ce{CH3OCH3}$, $\ce{C2H5OH}$, $\ce{NH2CHO}$ and $\ce{CH3CHO}$. The radial profile of COMs along two distinct directions (westward and southward from the centre) illustrates the drop in abundance away from the central source. Our models can successfully reproduce these observational features, with a highlight on the need for a higher cosmic-ray ionisation rate (e.g. at least 50-60 times higher than the accepted standard value of ISM at 0.2$\,$pc from the centre).

    \item[-] Comparison with Galactic massive clumps (G328.25, G335.58, G335.78): The observations of $\ce{CH3OH}$, $\ce{NH2CHO}$, $\ce{CH3OCH3}$, $\ce{C2H5OH}$ toward quiet massive infrared clumps (objects with bolometric luminosity less than $2\times 10^{4}L_{\odot}$), reported by \cite{2024A&A...686A.252B}, used the single-dish APEX telescope. The models that best reproduced these observations suggested temperature ranges that were broadly in line with the derivations from the observations. The need for an accurate treatment of the radiation and extinction in one dimension of the source structure is underscored by the way distance affects the data interpretations, as utilising a single-point model for hot core/hot corino could lead to an overestimation of COMs if they do not originate sufficiently near the central luminosity.

    \item[-] Comparison with ALMA observations towards Orion BN/KL: We considered only $\ce{CH3OH}$, $\ce{CH3OCH3}$, $\ce{HCOOH}$ and $\ce{C2H5OH}$ toward three well-known locations in the Orion BN/KL, including MF, EG, and ET peaks as reported in \cite{2018A&A...620L...6T}. Our model can nicely reproduce these observations, except for $\ce{C2H5OH}$ at the ET peak, showing that $\ce{HCOOH}$ emitting region is likely away from the centre, while other COMs are emitting near the centre.

    \item Our model cannot reproduce the high abundance of $\ce{C2H5OH}$ in SgrB2(N1) ($X(\ce{C2H5OH})>10^{-7}$) and in the peak of ethanol in Orion BN/KL ($X(\ce{C2H5OH})\geq3.5\times 10^{-8}$). The underlying reason is beyond the scope of this study. However, we showed that $X(\ce{C2H5OH})$ predicted from our 0D UCLCHEM is sensitive to the profile of the UV radiation field. It is also possible that our chemical network lacks of key reactions involving $\ce{C2H5OH}$.

\end{itemize}

A noteworthy limitation of our present model is the complete coupling of the dust temperature with the gas temperature, and the abundance estimated from our model is not average over the observed beams. Despite these limitations, our work showcased the UCLCHEM model's ability to interpret COMs detections across different astrophysical conditions, highlighting the need for a higher-dimensional model in comparison to the single-point model.

\begin{acknowledgements}
This work is financially supported by the advanced ERC grant ( ID: 833460, PI: Serena Viti) within the framework of MOlecules as Probes of the Physics of EXternal (MOPPEX) project. AC received financial support from the ERC Starting Grant “Chemtrip” (grant agreement No 949278). T.H. acknowledges the support from the main research project (No. 2025186902) from Korea Astronomy and Space Science Institute (KASI)
\end{acknowledgements}

\bibliographystyle{aa} 
\bibliography{bib}

\appendix
\section{Mean wavelength of radiation fields from the protostar and hot dust shell}
\begin{figure}
    \centering
    \includegraphics[width=0.8\linewidth]{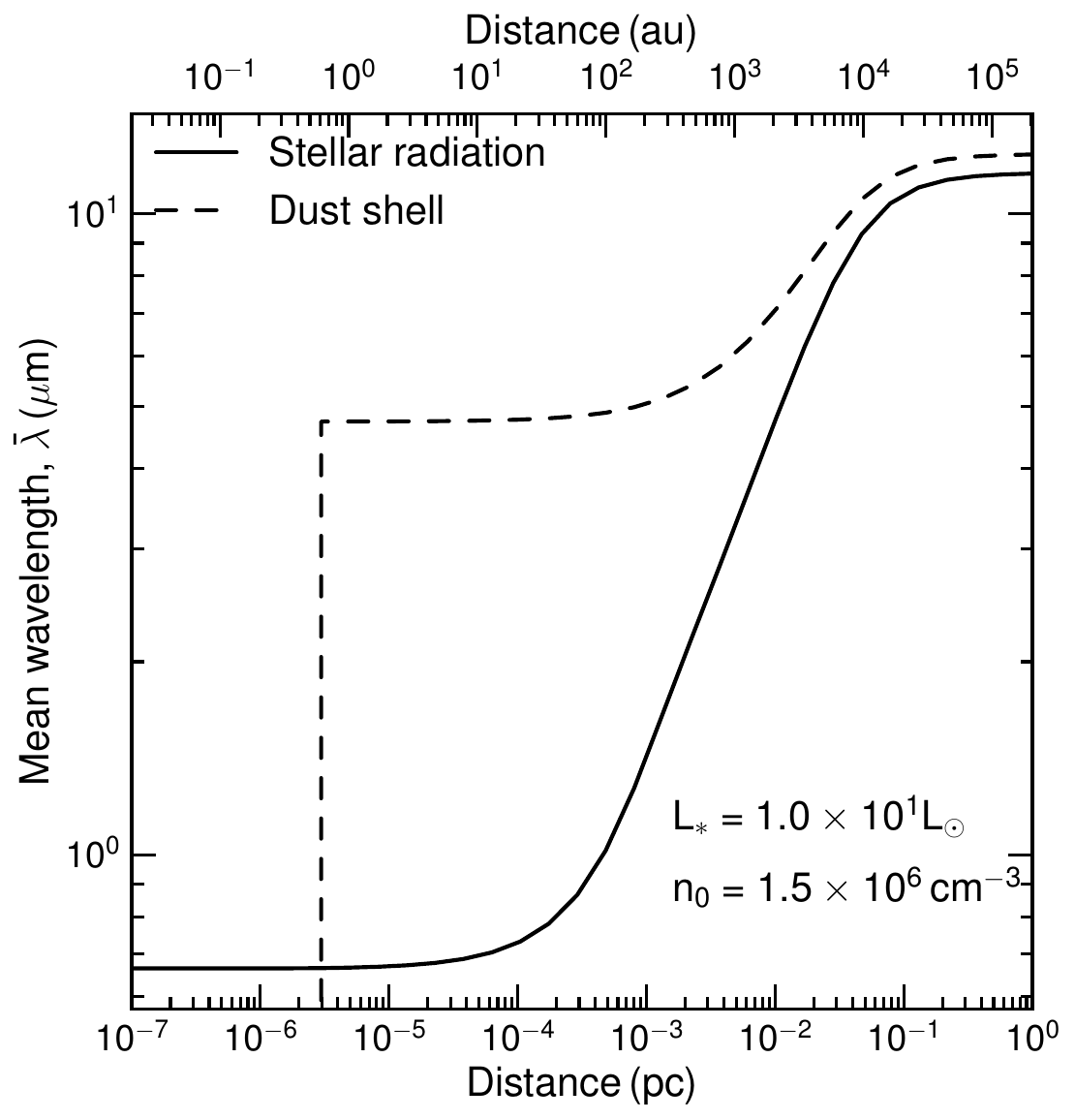}\\
    \includegraphics[width=0.8\linewidth]{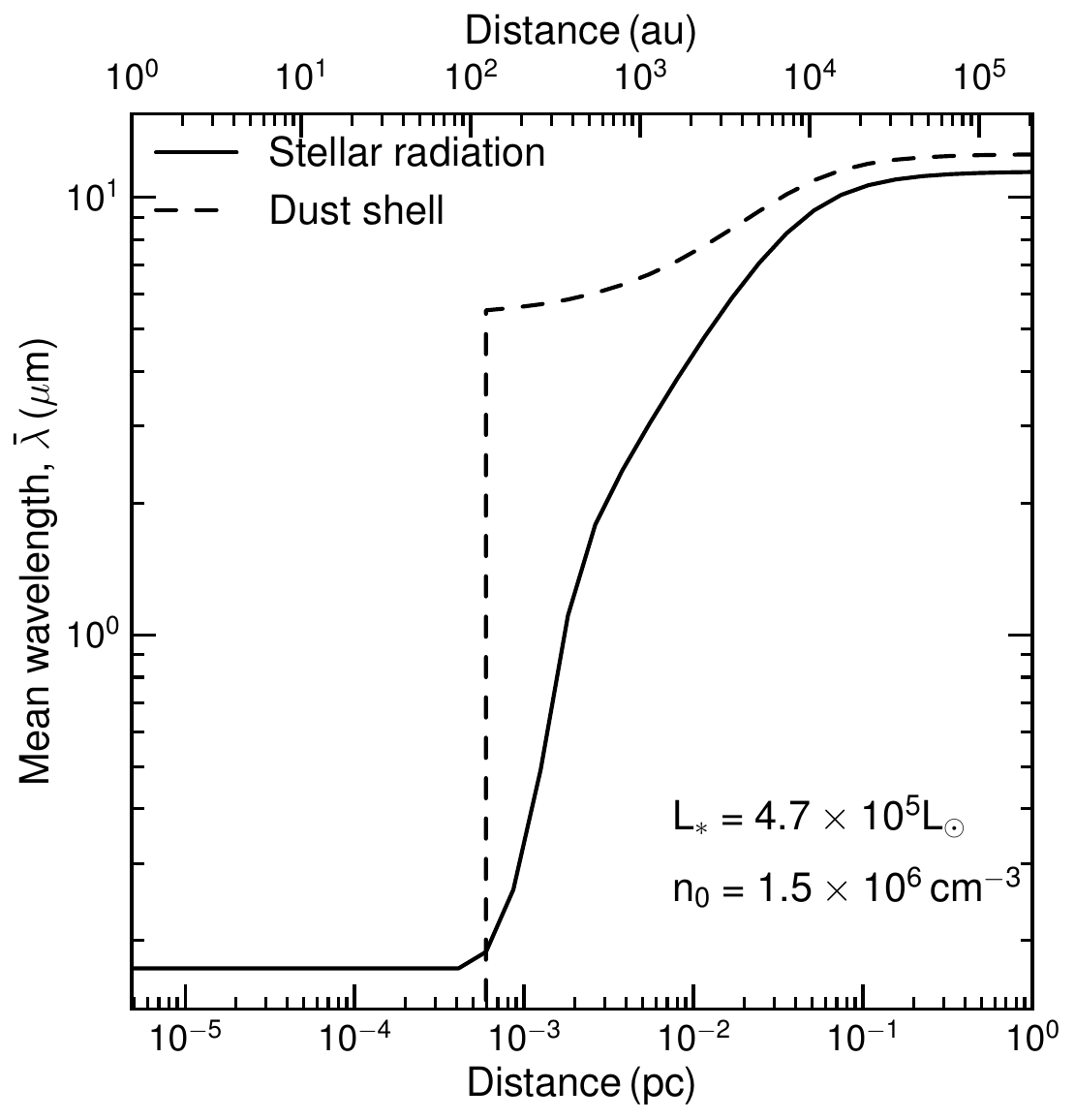}
    \caption{Decomposition of the radial profile of the radiation mean wavelength. The short-wavelength photons from the stellar radiations are effectively absorbed by the dust shell, allowing the longer-wavelength photons to penetrate to the outer region. The thermal radiation from the dust shell dominantly heats the outer region.}
    \label{fig:wave_decompose}
\end{figure}

Figure \ref{fig:wave_decompose} illustrates how the mean wavelength of the radiation field changes as it moves away from the source. UV photons released by massive protostars are mostly absorbed, allowing longer wavelength photons to propagate. In contrast, photons of longer wavelengths emitted by low-mass protostars can reach further into distant regions. Nevertheless, compared to the radiation field generated by the hot dust shell, the hot shell notably influences both the radiation and then the temperature within the envelope.   

\section{Radial profile of the cosmic-ray ionisation rate in SgrB2(N1) hot core}
\begin{figure}
    \centering
    \includegraphics[width=0.9\linewidth]{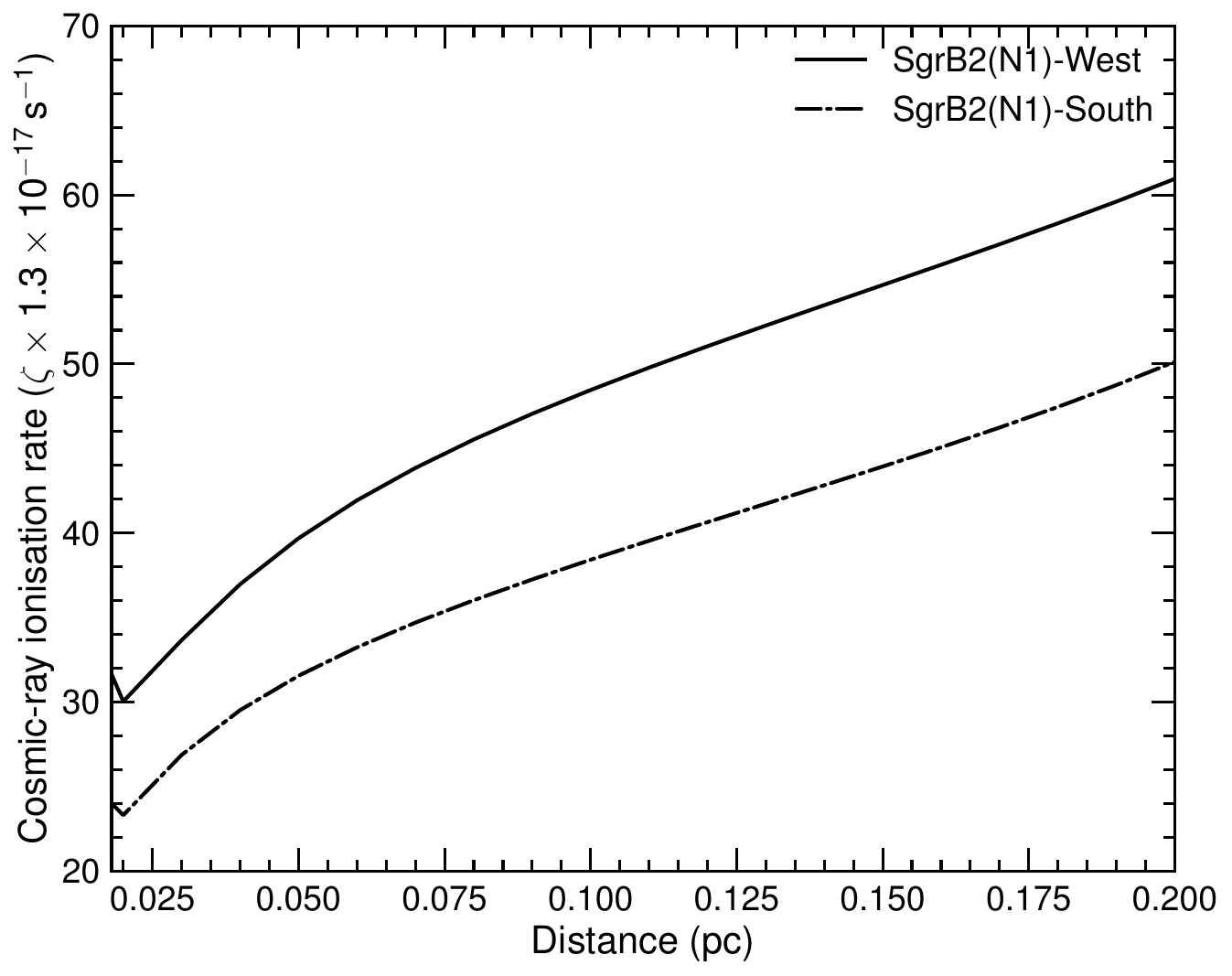}
    \caption{The radial profile of comis-ray ionisation rate in SgrB2(N1) hot core, induced from the best models shown in Figure \ref{fig:sgrb2n1}. The higher cosmic-rate ionisation rate is required in both directions, westward and southward.}
    \label{fig:zeta_sgrb2n1}
\end{figure}

Figure \ref{fig:zeta_sgrb2n1} shows the best profile of the cosmic-ray ionisation rate w.r.t distance from the central source, in the SrgB2(N1) hot core. This profile is based on the best $\zeta_{\rm scale}=30$ for the westward direction and $\zeta_{\rm scale}=20$ for the southward direction. At $0.2\,$pc away from the centre, a higher rate is seen with $\zeta \simeq 50-60\times \zeta_{\rm ISM}$. This high number is consistent with \cite{2019A&A...628...A27}.

\section{Effect of grain size and UV radiation field}
\begin{figure*}[!ht]
    \centering
    \includegraphics[width=0.95\linewidth]{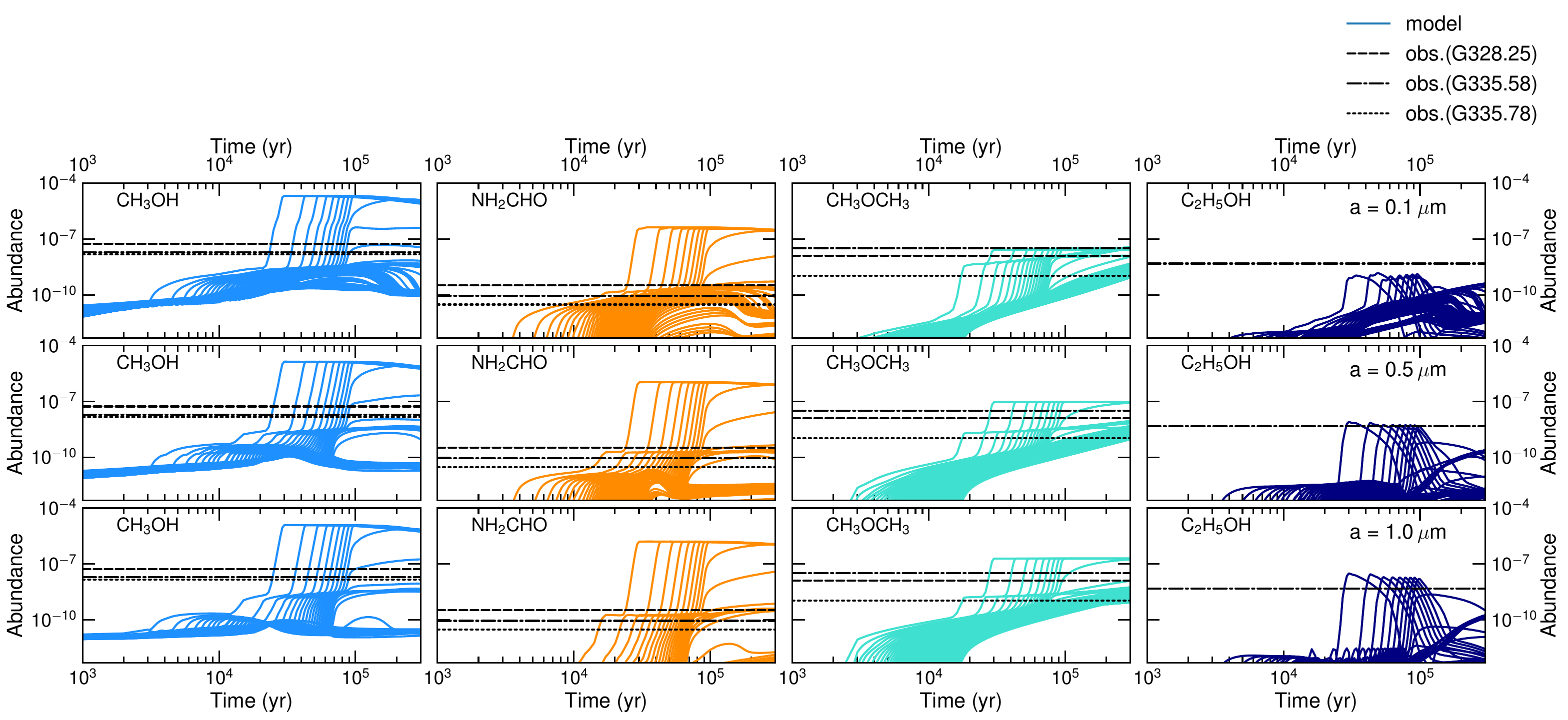}
    \caption{The time-dependent behavior of COMs (colour lines, different lines corresponding to different distance as in Figure \ref{fig:Xch3oh}) is compared with measurements (horizontal black lines) taken from three infrared quiet galactic massive clumps (G328.25, G335.58, and G335.78) with $L=10^{4}\,L_{\odot}$. To explain better all the COMs (including C$_{2}$H$_{5}$OH) abundance reported in G335.58, the average grain size is shown to be larger than the typical grain size of $0.1\,\mu$m in diffuse ISM.}
    \label{fig:Galactic_massive_clumps}
\end{figure*}

Figure \ref{fig:Galactic_massive_clumps} illustrates the time variation in the abundances for a different location in the cloud (indicated by coloured lines) and grain sizes. Closer to the central light source, a rapid increase in abundance is observed as a result of elevated temperatures, compared to locations further in the cloud. These abundances are then compared with the above-mentioned data (black horizontal lines). Thus, to explain the observed data, $\ce{CH3OH}$ and $\ce{NH2CHO}$ must be emitted further away from the centre, while $\ce{CH3OCH3}$ and $\ce{C2H5OH}$ originate nearer to the centre. Furthermore, when the grain size is $0.1\,\mu$m, our model underestimates the abundance of \ce{C2H5OH}. By increasing the grain size to $0.5\,\mu$m and $1\,\mu$m, we find that to match the observed abundance of all COMs (including $\ce{C2H5OH}$), the average grain size must be around $0.5\,\mu$m. This value is larger than the typical grain size in the diffuse ISM.

\begin{figure*}
    \centering
    \includegraphics[width=0.85\linewidth]{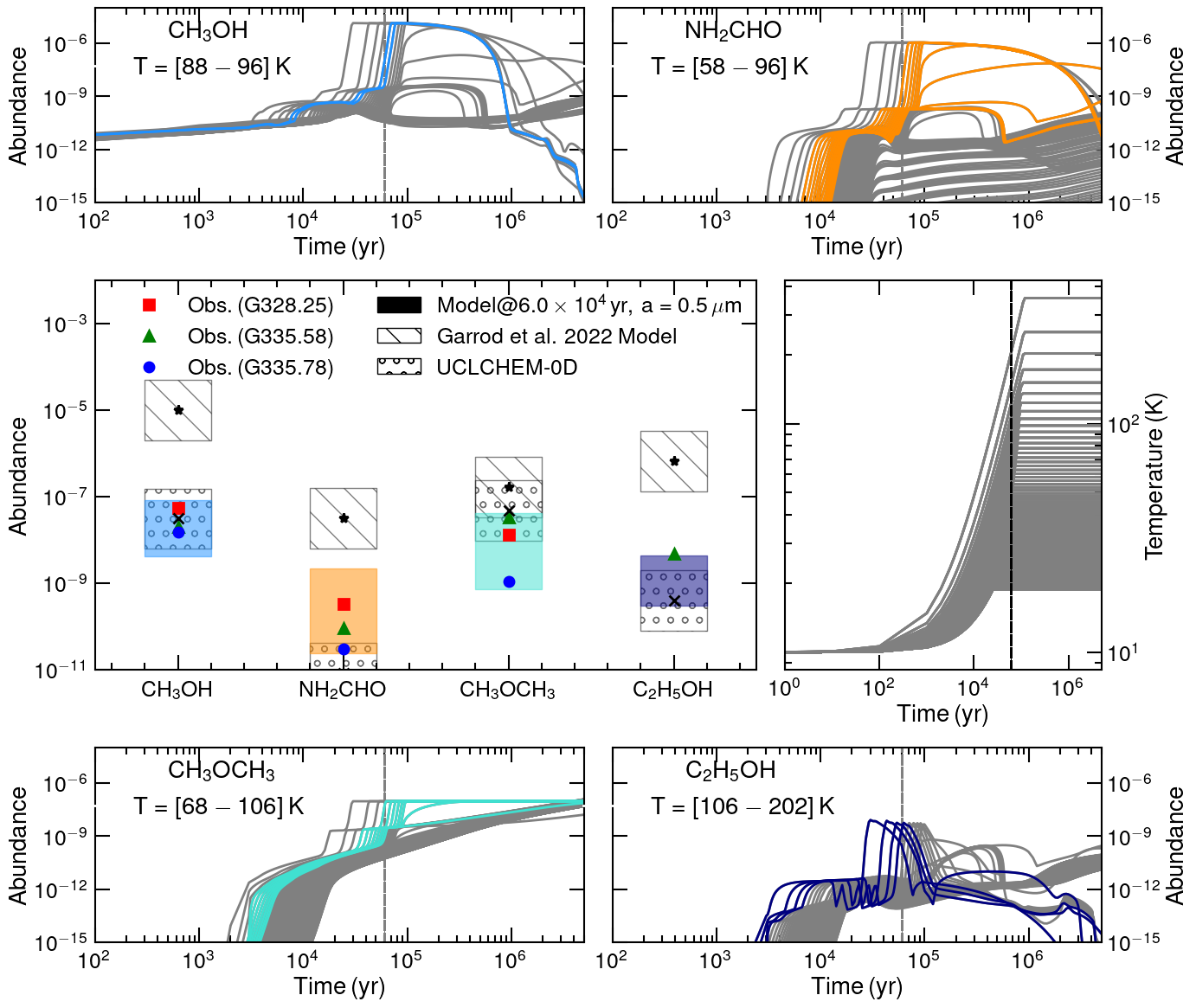}
    \caption{Similar to Figure \ref{fig:Galactic_massive_clumps_shaded} but the UCLCHEM-0D models for $G_{0}=10$. The COMs are less abundant for higher values of the UV radiation field.}
    \label{fig:Galactic_massive_clumps_shaded_G0=10}
\end{figure*}

Figure \ref{fig:Galactic_massive_clumps_shaded_G0=10} shows the abundance of COMs at $G_{0}=10$, as determined by the UCLCHEM-0D model. In contrast to the scenario of $G_{0}=1$, the COMs exhibit significantly reduced abundances. Consequently, employing the 0D model may lead to a deficiency of UV-related processes in protostars due to the rapid attenuation of external UV radiation. Conversely, the 1D model highlights the significant influence of the central source's internal UV field.

\end{document}